\documentclass[prb,twocolumn,superscriptaddress,footinbib]{revtex4-1} 
\usepackage{longtable}
\usepackage{morefloats}
\usepackage[dvips]{graphicx}
\usepackage{color}
\usepackage{epsfig,graphicx,amsfonts,amsbsy}
\usepackage{amsmath,amsfonts,amsthm,amssymb}
\usepackage{appendix}
\usepackage{makeidx}
\usepackage{url}
\usepackage{verbatim}
\usepackage[bookmarksnumbered,pdfpagelabels=true,plainpages=false,colorlinks=true,linkcolor=blue,citecolor=blue,urlcolor=blue]{hyperref}
\usepackage[normalem]{ulem}
\usepackage[rightcaption]{sidecap}
\usepackage{array}
\usepackage{breakurl}
\usepackage{xcolor,cancel}
\usepackage{multirow}
\usepackage{tabularx}
\usepackage{braket} 
\usepackage{dsfont}
\usepackage{amsmath}
\usepackage{ulem}
\usepackage{bbold}
\usepackage{etoolbox}

\newcommand{\pt}{\partial}

\newcommand{\mc}{\mathcal}

\newcommand\norm[1]{\left\lVert#1\right\rVert}
\newcommand{\ccr}{c_{\scalebox{0.6}{cr}}}

\begin{document}

\title{Energy Transfer in Random-Matrix ensembles of Floquet Hamiltonians}

\author{Christina Psaroudaki}
\affiliation{Laboratoire de Physique de l’\'{E}cole normale sup\'{e}rieure, ENS, Universit\'{e} PSL, CNRS, Sorbonne Universit\'{e}, Universit\'{e} de Paris, F-75005 Paris, France}
\author{Gil Refael}
\affiliation{Department of Physics and Institute for Quantum Information and Matter, California Institute of Technology, Pasadena, CA 91125, USA}
\affiliation{Walter Burke Institute for Theoretical Physics, California Institute of Technology, Pasadena, CA 91125, USA}

\date{\today}
\begin{abstract}
We explore the statistical properties of energy transfer in ensembles of doubly-driven Random-Matrix Floquet Hamiltonians, based on universal symmetry arguments. The energy pumping efficiency distribution $P(\bar{E})$ is associated with the Hamiltonian parameter ensemble and the eigenvalue statistics of the Floquet operator. For specific Hamiltonian ensembles, $P(\bar{E})$ undergoes a transition which cannot be associated with a symmetry breaking of the instantaneous Hamiltonian. The Floquet eigenvalue spacing distribution indicates the considered ensembles constitute generic nonintegrable Hamiltonian families. As a step towards Hamiltonian engineering, we develop a machine-learning classifier to understand the relative parameter importance in resulting high conversion efficiency. We propose Random Floquet Hamiltonians as a general framework to investigate frequency conversion effects in a new class of generic dynamical processes beyond adiabatic pumps. 
\end{abstract}

\maketitle
\section{Introduction}

Periodic driving of a quantum system is a versatile tool for its coherent control, allowing one to engineer quantum phases of matter with various applications \cite{RevModPhys.89.011004}. It opens up the possibility to artificially realize exotic topological systems \cite{Lindner2011,PhysRevX.6.021013,PhysRevLett.120.150601,PhysRevX.3.031005,PhysRevX.6.041001,PhysRevB.93.201103,PhysRevB.82.235114}, many of which have no static analog \cite{PhysRevLett.106.220402}. Among them, a class of double-drive Hamiltonians displays quantized adiabatic pumping of energy between the two drives \cite{PhysRevX.7.041008,PhysRevB.97.134303,PhysRevB.99.094311}, in close analogy with the Thouless topological charge pumping \cite{PhysRevB.27.6083} and its inverse effect in adiabatic quantum motors \cite{PhysRevLett.111.060802,Switkes1905}. 

Our contribution to the collection in honor of Emmanuel Rashba focuses on energy pumping in doubly-driven systems.   Indeed, energy pumping between multiple drives, is analogous to the anomalous Hall effect in spin orbit coupled bands. With this, the seminal work of Rashba on spin-orbit effects in solids\cite{Rashba2015SymmetryOE}
 finds an application in the synthetic-dimension picture of multiply driven systems.
 
Energy pumping between multiple drives could be an crucial element for quantum machines and amplifiers at the terahertz regime. Quantized energy flow between two incommensurate drives is predicted in temporal analogs of two-dimensional topological insulators \cite{PhysRevX.7.041008,PhysRevB.99.064306}.  
Nevertheless, quantized pumping emerges as long as the system is in the near-adiabatic limit, during which any instantaneous bulk gap is maintained, and is restricted by the topology of the relevant band \cite{RevModPhys.82.1959}. Quantized energy transfer has been predicted only for specific double-drive topological models inside the model's topological phase and for irrationally-related drive frequencies \cite{PhysRevX.7.041008}. Energy flow outside this relatively small part of the parameter space is practically unexplored and is not expected to remain robust to nonadiabatic driving conditions \cite{PhysRevLett.120.106601}. In this limit, the regime of harmonic frequency ratios is particularly interesting as it allows an energy conversion rate exceeding the quantized value in both the topological and trivial class \cite{PhysRevX.7.041008} and a sustained response in the nonadiabatic regime \cite{PSAROUDAKI2021168553}.

Here, we propose Random Floquet Hamiltonians as a general framework to investigate frequency conversion effects in a relatively large parameter space and a powerful tool to explore a new class of generic dynamical processes beyond adiabatic pumps. Since Wigner’s original proposal on the use of random matrices to describe properties of highly excited nuclear levels in complex nuclei \cite{Wigner57,Porter1965,Wigner67,GUHR1998189}, Random Matrix Theory (RMT) has been applied in a variety of physical problems, including quantum transport \cite{RevModPhys.69.731,RevModPhys.72.895} and quantum chaotic systems \cite{PhysRevLett.52.1}. Inspired by the universality of RMT, we study the statistical properties of the energy-pumping effect for an ensemble of doubly-driven Random Floquet Hamiltonians. Of primary interest is the characterization of the energy pumping efficiency distribution and its relation to Hamiltonian distributions and Floquet level statistics. We use the basic properties of RMT as a standard diagnostic tool of generic nonintegrable ensembles.

From an analysis of various instantaneous Hamiltonian symmetries, it follows that the energy pumping efficiency distribution $P(\bar{E})$ depends on the Hamiltonian parameter ensemble. For a Gaussian Hamiltonian distribution, the energy pumping has no linear correlation to the Hamiltonian norm. Remarkably, for a spherical Hamiltonian ensemble and a Hamiltonian ensemble with complex parameters, $P(\bar{E})$ undergoes a transition that cannot be associated with a symmetry breaking of the instantaneous Hamiltonian. The Floquet spacing statistics exhibit a linear (quadratic) level repulsion at small spacings for Hamiltonians with real (complex) parameters, and its form indicates the considered ensemble constitutes a generic Hamiltonian family. In all cases we considered, we find the universal behavior $P(\bar{E}) = f(\bar{E}/\beta)$ with $\beta \propto \sigma^4$, and $\sigma$ being a scale parameter specific to the distribution. As a step towards Hamiltonian engineering targeting high conversion efficiency, we develop a machine-learning model classifier to extract the importance of each Hamiltonian parameter, applied to a class of random temporal topological models.  Our results can be implemented in a variety of double-drive two-level systems, including a spin-1/2 \cite{PhysRevX.7.041008}, single-qubit systems \cite{PhysRevLett.125.160505, PhysRevLett.126.163602} and non-interacting atoms trapped in an optical cavity \cite{PhysRevB.99.094311}.

The structure of the paper is as follows. In Sec.~\ref{sec:FreqConv} we introduce the energy pumping in a family of two-frequency Hamiltonians. In Sec.~\ref{sec:RandomFloquet} we numerically calculate the energy pumping efficiency, while in Sec.~\ref{sec:MachineLearning} we develop a machine-learning classifier. A discussion on analytical bounds is included in Sec.~\ref{sec:AnalyticalBounds}. Our main conclusions are summarized in Sec.~\ref{sec:Discussion}, while some technical details are deferred to three Appendices.

\section{Frequency Conversion} \label{sec:FreqConv}

Our analysis begins by considering a family of two-frequency Hermitian $2\times 2$ Hamiltonians
 \begin{align}
H(t)=F_0 + H_1(\omega_1 t +\phi_1)+H_2(\omega_2 t +\phi_2) \,, \label{eq:Hamiltonian}
\end{align} 
with $H_i$ being periodic function of time with frequency $\omega_i$ and phase $\phi_i$. The main quantity of interest is the  integrated power absorbed (or spent) by the drive $i=1,2$, 
\begin{align}
E_i (t)=\int_0^{t} dt' \langle \psi(t') \vert \frac{d H_i(t')}{dt'} \vert \psi (t') \rangle \,, \label{eq:EnergyBHZ} 
\end{align}
where $\vert \psi (t) \rangle = U(t) \vert \psi_0 \rangle$ is the instantaneous eigenstate, $U(t) = \mathcal{T} \exp[-i \int_0^{t} H(t') dt'] $ is the time evolution operator with $\mathcal{T}$ the time-ordering operator, and $\vert \psi_0 \rangle $ the initial state. Each $E_i(t)$ depends linearly on time, although the rates of work performed by the two sources add up approximately to zero, $\bar{E}= \lim_{t\rightarrow \infty}E_1(t)/t = -\lim_{t\rightarrow \infty}E_2(t)/t$. We aim to characterize the frequency conversion efficiency $\bar{E}$ for generic Floquet Hamiltonians. 

In the special case of the temporal analog of the topological Bernevig-Hughes-Zhang (BHZ) model \cite{Bernevig1757} the Hall response translates to a quantized pumping of energy $\bar{E}= \omega_1 \omega_2/2\pi=E_Q$ which emerges in the adiabatic limit $\eta \gg \omega_i$, with $\eta$ the driving amplitude, rationally independent frequencies $\omega_1/\omega_2 \equiv \gamma$ with $\gamma \notin \mathbb{Q}$, and topological regime \cite{PhysRevX.7.041008}. Rationally-related frequencies $\omega_1/\omega_2 \equiv q/p$ for $p,q  \in \mathbb{Z}$, exhibit a sustained response which could exceed the quantized value $E_Q$ in both the topological and trivial regime \cite{PhysRevX.7.041008}, as well as in the nonadiabatic limit $\eta \ll \omega_i$ \cite{PSAROUDAKI2021168553}. Details of the energy pumping for the temporal BHZ model are summarized in the Appendix \ref{app:QuantinzedPumping}.
\begin{figure}[t]
	\centering
	\includegraphics[width=1\linewidth]{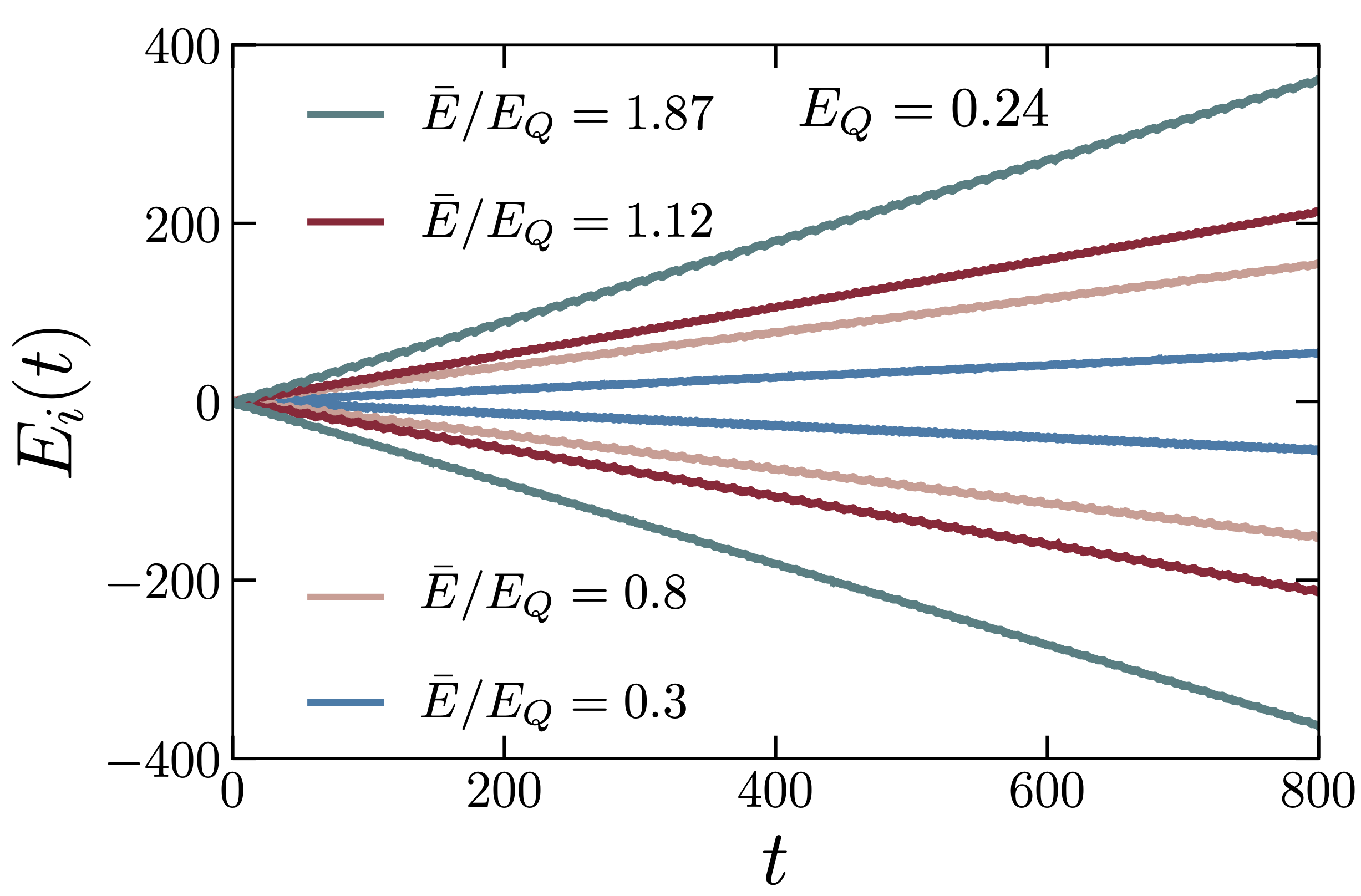}
	\caption{Energy flow $E_i(t)$ between the two drives of rationally-related frequencies $\omega_2/\omega_1=3/2$ with Hamiltonian parameters chosen from a Gaussian ensemble with $c=1$ and $\sigma=1$. Four distinct Floquet systems with approximately the same driving amplitude $\bar{F}=( \norm{F_0}+\norm{F_1}+\norm{F_2})/3\approx 0.7$, display a distinct dynamical behavior. }
	\label{Fig:PE_t}
\end{figure}

\begin{figure}[b]
	\centering
	\includegraphics[width=1\linewidth]{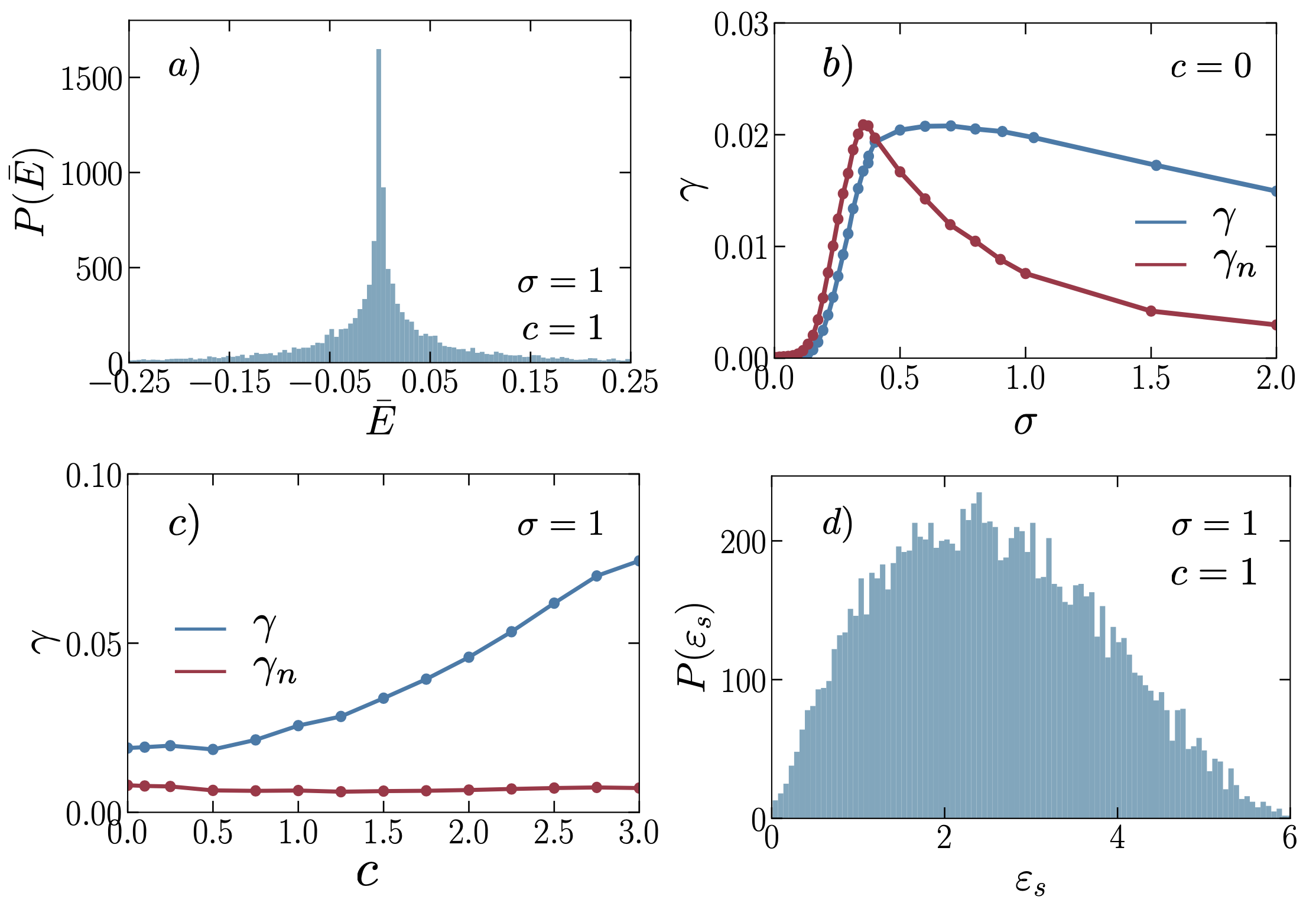}
	\caption{Energy pumping distribution $P(\bar{E})$ for the real Gaussian ensemble (Sec. \ref{G-sec}) and a sampling of $N=12000$ realizations. (a) $P(\bar{E})$ for (offset) $c=1$ and (width) $\sigma=1$, approximated by a Lorentzian curve $P(\bar{E}) \propto [1+\bar{E}^2/\gamma^2]^{-1}$, with $\gamma=0.024$. (b) The dependence of $\gamma$ and $\gamma_n$ on $\sigma$ for $c=0$, with $\gamma_n$ the scale parameter of the  distribution of the pumping rate normalized by amplitude, $P(\bar{E}/\norm{F_i})$. (c) The dependence of $\gamma$ and $\gamma_n$ on $c$ for $\sigma=1$. (d) Distribution of the nearest neighbor Floquet spacing $P(\varepsilon_s)$, indicating level repulsion.}
	\label{Fig:DN}
\end{figure}

The goal of the present work is to characterize generic frequency conversion dynamical processes generated by random Floquet Hamiltonians. In the most general case, adiabatic cycles where there is always an energy gap in the instantaneous  spectrum should not be expected. It thus appears promising to focus on 
 rationally-related frequencies such that the system is strictly periodic with a period equal to $T \equiv 2\pi/p\omega_1 = 2\pi/q \omega_2$. To maximize the energy transfer, we consider Floquet eigenstate initialization with $\vert \psi_0  \rangle$ the lowest of the two Floquet eigenstates $U(T)\vert  \psi_n \rangle =e^{-i \varepsilon_n T} \vert  \psi_n \rangle$. $U(T)$ is the Floquet single-period evolution operator \cite{Floquet1883}. 
 
When the two frequencies form a rational fraction, a further simplification of the energy formula can be obtained. For a general multi-driven Floquet problem it holds
\begin{align}
&E_i (t)=\int_0^{t} dt' \langle \psi(t') \vert \omega_i\frac{\partial H(t') }{\partial \phi_i} \vert \psi(t') \rangle \nonumber \\
&=\omega_i\int_0^{t} dt' \langle \psi(0) \vert U^{\dagger}(t') \frac{\partial H(t') }{\partial \phi_i} U(t')\vert \psi(0)\rangle \nonumber \\
&=i\omega_i\langle \psi(0)  \vert U^{\dagger}(t) \frac{\partial U(t) }{\partial \phi_i}\vert \psi(0) \rangle  \nonumber \\
&=i\omega_i\langle \psi(0)  \vert  \frac{\partial \log U(t) }{\partial \phi_i}\vert \psi(0) \rangle=i\omega_i\frac{\partial }{\partial \phi_i}\langle \psi(0)  \vert   \log U(t)\vert \psi(0) \rangle \,.
\label{simple}
\end{align}

Thus, the work done is related directly to the dependence of the trace log of the evolution operator. In the above we used 
\[
\frac{\partial U(t)}{\partial \phi_i}=\int\limits_0^t dt' U(t',t) \left(-i  \frac{\partial H(t') }{\partial \phi_i}\right)U(0,t')
\]
with $U(t_1,t_2)$ the propagator between the times $t_1$ and $t_2$. Next, let us concentrate on the rational-fraction case, where the system has a time-periodic Hamiltonian with period $T$. Also, we assume that the system is initiated into a Floquet eigenstate $\ket{\psi_{\varepsilon}}$ with Floquet eigenenergy $\varepsilon$. Following from Eq. (\ref{simple})
\begin{align}
&E_i (T)=i\omega_i\langle \psi_{\varepsilon} \vert  \frac{\partial \log U(T) }{\partial \phi_i}\vert \psi_{\varepsilon} \rangle  \nonumber \\
&=i\omega_i\langle \psi_{\varepsilon} \vert  \frac{\partial}{\partial \phi_i}\left(\log U(T)\vert \psi_{\varepsilon}\right) \rangle-
i\omega_i\langle \psi_{\varepsilon} \vert \log U(T) \frac{\partial \vert \psi_{\varepsilon} \rangle}{\partial \phi_i}  \nonumber \\
&=T\omega_i \frac{\partial \varepsilon}{\partial \phi_i}
\end{align}
where T is the period of the combined drive. The average power exchanged by the drives is therefore
\begin{equation}
    \bar{E}_i=\omega_i \frac{\partial \varepsilon}{\partial \phi_i} \,.
\end{equation}

\section{Random Floquet Ensembles} \label{sec:RandomFloquet}

\subsection{Gaussian Ensemble} 
\label{G-sec}

With these preliminary remarks, we now begin our analysis by considering the model of Eq.~\eqref{eq:Hamiltonian} with
\begin{equation}
\begin{array}{c}
H_0+H_1 = F_0+F_1 \cos(\omega_1 t +\phi_1)\vspace{0.2cm}\\
H_2 =F_2 \cos(\omega_2 t +\phi_2),\vspace{0.2cm}\\
F_i = h_0^{i} \mathbb{1} + h_x^i \sigma_x + h_y^i \sigma_y+h_z^i \sigma_z.
\label{hG}
\end{array}
\end{equation}
Here $\sigma_i$ are the Pauli matrices. All $h_j^i$ are chosen from a Gaussian distribution 
\begin{equation}
P(h_i^j) = e^{-(h_i^j -c)^2/2 \sigma^2}.\label{distG}
\end{equation}
We use $\omega_1/\omega_2 =2/3$, $\omega_2=1$ and $\phi_2 =0=\phi_1$ throughout, and sample over $N=12000$ realizations of $H(t)$. In Fig.~\ref{Fig:PE_t} we depict the energy transfer between the two drives $E_i(t)$ for four distinct Floquet systems chosen from a Gaussian distribution with $c=1$ and $\sigma=1$, but with approximately the same amplitude $\bar{F}=( \norm{F_0}+\norm{F_1}+\norm{F_2}) /3\approx 0.7$. The resulting energy pumping efficiency varies substantially between the  different realizations and could exceed the quantized value $E_Q$, indicating that the various Floquet states exhibit distinct dynamical behavior. 

\begin{figure}[t]
	\centering
	\includegraphics[width=1\linewidth]{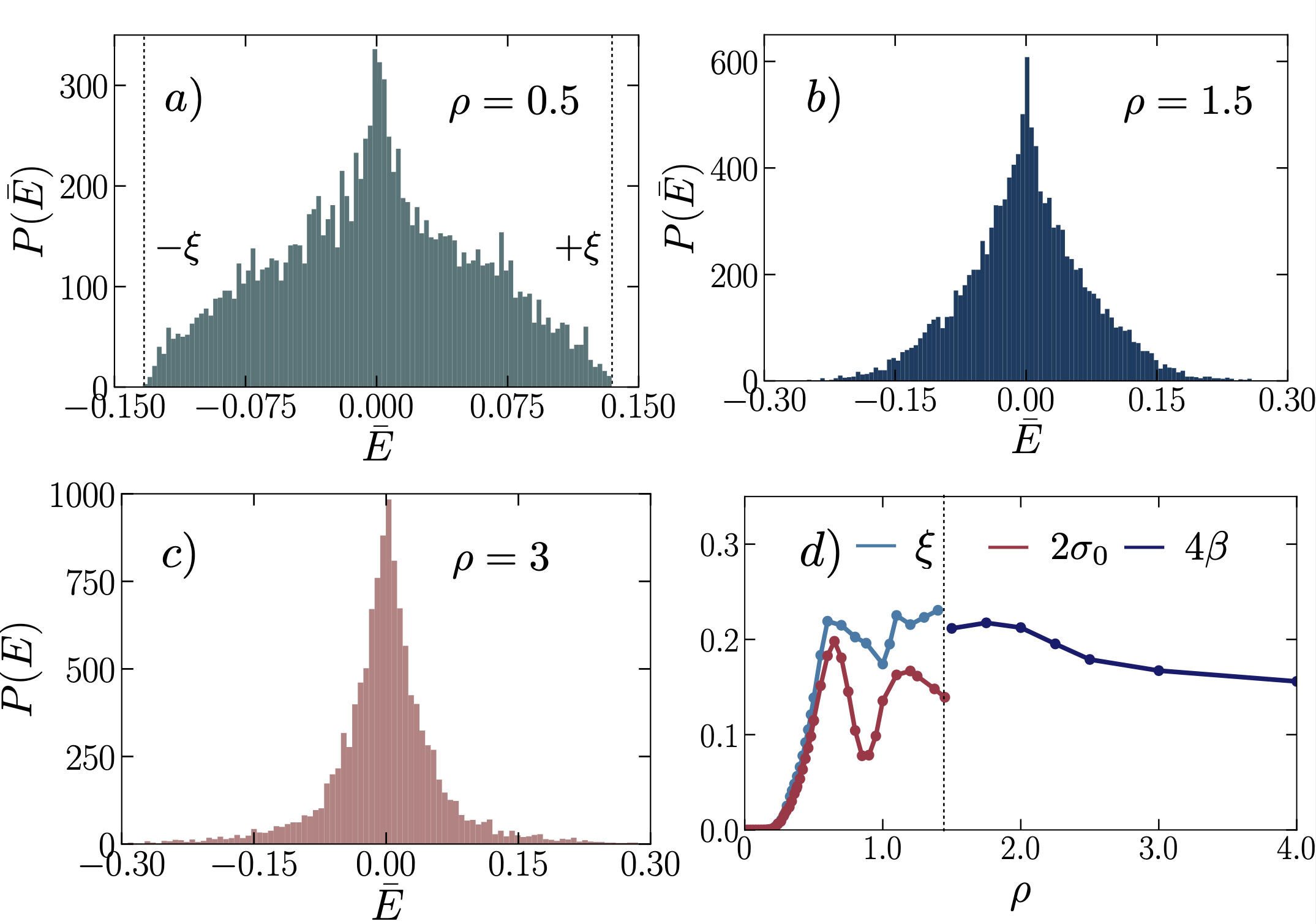}
	\caption{Energy pumping distribution $P(\bar{E})$ for the spherical ensemble of Sec. \ref{SE-sec} with distribution norm $\rho$ and $N=12000$ realizations. (a) For $\rho =0.5<\rho_c $, $P(\bar{E})$ is approximated by a symmetric triangular distribution which terminates at $\bar{E}=\pm\xi$. (b)-(c) Above $\rho_c$, $P(\bar{E}) \propto e^{-\vert \bar{E} \vert/\beta}$ is described by a Laplacian curve. (d) Parameters $\xi$, $\beta$ and standard deviation $\sigma_0$ as a function of $\rho$, with a transition at $\rho_c \approx 1.5$.}
	\label{Fig:UN}
\end{figure}

In Fig.~\ref{Fig:DN} we summarize the statistical properties of the energy pumping efficiency distribution $P(\bar{E})$.  For any value of $\sigma$ and $c$, $P(\bar{E})$ is well approximated by a Lorentzian curve 
\begin{equation}
P(\bar{E}) \propto [1+\bar{E}^2/\gamma^2]^{-1}.
\end{equation}
In Fig.~\ref{Fig:DN}-a) we depict $P(\bar{E})$ for $c=1$ and $\sigma=1$ described by $\gamma=0.024$. Since we are interested in the nonadiabatic limit $c \ll \omega_i$, we study the energy pumping efficiency for $c=0$, with the relevant scale now given by $\sigma$. The scale parameter $\gamma$ plotted in Fig.~\ref{Fig:DN}-b) grows as $\gamma \propto \sigma^4$ for $\sigma \ll 1$ and decreases for $\sigma \gtrsim 0.5$. The normalized distribution $P(\bar{E}/\norm{F_i})$ is described by $\gamma_n$ with a similar behavior. For a given $\sigma=1$, $\gamma$ is an increasing function of $c$ and $\gamma_n$ a weakly-dependent [see Fig.~\ref{Fig:DN}-c)], indicating that the Hamiltonian strength dominates the pumping strength at the nonadiabatic regime. The considered model belongs in the trivial dynamical class $C=0$,  where $C$ is the Chern number associated with the instantaneous ground state band.  In the Appendix \ref{app:berrycurvature} we discuss the geometric aspects of the energy pumping effect encoded in the Berry curvature of the quasienergy state for various Hamiltonian realizations,  and provide analytical expressions for $C$.
 
As the universality of transport properties is related to the level statistics and spectral correlations \cite{RevModPhys.69.731,Porter1965}, it is natural to study the level statistics of the Floquet operator $U(T)$. We focus on the nearest neighbor spacing distribution $P(\varepsilon_s)$ between two adjacent ordered levels, $\varepsilon_s = (\varepsilon_2-\varepsilon_1)/\langle \varepsilon_s \rangle$. Level statistics have been used in random Floquet systems to understand the statistics of Floquet operators \cite{Dietz1994}, Floquet thermalization \cite{PhysRevB.93.104203}, and disorder in driven topological phases \cite{PhysRevLett.121.126803}. In Fig.~\ref{Fig:DN}-(d) we depict $P(\varepsilon_s)$ for $\sigma=1$ and $c=1$, well approximated by $P(\varepsilon_s)\propto \varepsilon_s e^{-b \varepsilon_s^2}$, with $b=0.12$ and a linear level repulsion at small spacings. The form of $P(\varepsilon_s)$ resembles the spacing distribution of a Gaussian orthogonal ensemble with $b=\pi/4$ \cite{GUHR1998189}, and indicates that the considered ensemble constitutes a generic Hamiltonian family. The Floquet operator can be in a different random matrix class to the instantaneous Hamiltonian. Finally, no linear relationship can be established between $\bar{E}$ and both the instantaneous and time-averaged Hamiltonian norm $\bar{H}_0= \norm{H(t=0)}/c_0$ and $\bar{H}=\norm{F_0}/c_0$, with $c_0=3 \norm{\mathbb{1}+\sum_i \sigma_i}$ a normalization constant. The two datasets are characterized by an almost vanishing correlation coefficient $r=0.05$ (see Appendix \ref{app:QuantinzedPumping} Fig.~\ref{Fig:ScatterPlot} for the dependence of $\bar{E}$ on either $\bar{H}_0$ and $\bar{H}$ and Eq.~\eqref{seq:PearsonsCoeff} for the definition of $r$). 
\begin{figure}[t]
	\centering
	\includegraphics[width=1\linewidth]{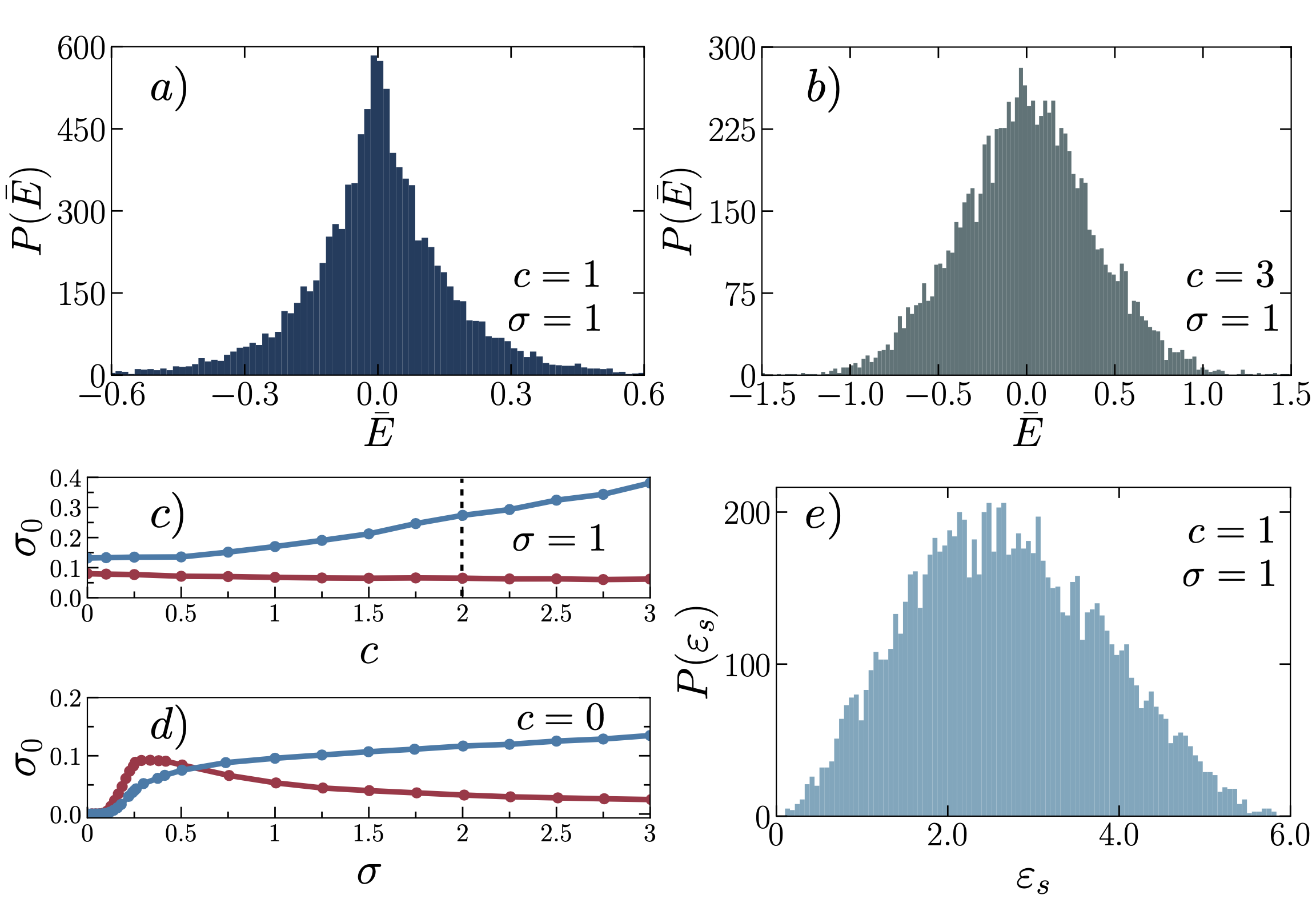}
	\caption{Energy pumping distribution $P(\bar{E})$ for the complex Gaussian Hamiltonian ensemble. These results represent sampling of $N=10^4$ realizations. a) For offset $c=1<\ccr$, $P(\bar{E})$ is described by a Laplacian curve, while b) for $c=3>\ccr$ by a Gaussian, with $\ccr=2$. c) Standard deviation $\sigma_0$ of $P(\bar{E})$ (blue line) and $\sigma_n$ of the distribution of nomrmalized pumping, $P(\bar{E}/\norm{F_i})$ (red line) as a function of $c$ for $\sigma=1$. d) Standard deviation $\sigma_0$ and $\sigma_n$ as a function of $\sigma$ for $c=0$. e) Nearest-neighbor Floquet spacing distribution $P(\varepsilon_s)$ for $\sigma=1$ and $c=1$, indicating no sign of integrability}.
	\label{Fig:CP}
\end{figure}

\subsection{Spherical Ensemble} \label{SE-sec}

We now consider the model of Eq.~\eqref{eq:Hamiltonian} with \begin{equation}
\begin{array}{c}
H_0+H_1(t)= F_0+(\mathbf{G_1}\cdot \boldsymbol{\sigma}) \cos(\omega_1 t +\phi_1),\vspace{2mm}\\
H_2(t)=( \mathbf{G_2}\cdot \boldsymbol{\sigma}) \cos(\omega_2 t +\phi_2), \vspace{2mm}\\
F_0=\mathbf{G}_3 \cdot \boldsymbol{\sigma},\vspace{2mm}\\
\mathbf{G}_i = \rho( \sqrt{1-\chi_i^2}\cos\theta_i ,\sqrt{1-\chi_i^2}\sin\theta_i, \chi_i ),
\end{array}
\end{equation}
where $\chi_i$ is chosen from a uniform distribution on the interval $[-1,1]$ and $\theta_i$ on the interval $[0,2\pi]$, providing Haar-measure sampling. Configurations obtained by an SU(2) transformation of vectors $\mathbf{G}_i$, result in the same energy pumping efficiency. The resulting behavior is summarized in Fig.~\ref{Fig:UN} for different values of the distribution norm $\rho$, $\omega_1/\omega_2=2/3$, $\omega_2=1$ and $\phi_1=0=\phi_2$. Quite surprisingly, there is a critical value $\rho_c \approx 1.5$, above which $P(\bar{E})$ changes from a symmetric triangular distribution with support at $|\bar{E}|< \xi$:
\begin{equation}
P(\bar{E})\approx \frac{1}{\xi^2}(\bar{E}-\xi)\,\,\,\,\, [\rho<1.5]    
\end{equation}
to a distribution approximated by a Laplacian:
\begin{equation}
P(\bar{E}) \approx \frac{1}{\beta} e^{-\vert \bar{E} \vert/\beta}\,\,\,\,\, [\rho>1.5].   .
\end{equation}
This behavior cannot be associated with an instantaneous Hamiltonian symmetry breaking. For $\rho\ll 1$ we find the scaling $\sigma_0 \propto \rho^4$, with $\sigma_0$ the standard deviation of the triangular distribution. Nearest neighbor spacing distribution $P(\varepsilon_s)$ is approximated by a curve of the form $P(\varepsilon_s)\propto \varepsilon_s e^{-b \varepsilon_s^2}$ with $b=0.16$ and a linear level repulsion at small spacings (see Fig.~\ref{Fig:UNC1} of Appendix \ref{app:QuantinzedPumping}). For all model realizations, it holds that $C=0$.

\subsection{Complex Gaussian Ensemble} \label{CG-sec}

To complete the description, we now turn to generalized two-frequency models with complex parameters of the form 
\begin{equation}
\begin{array}{c}
H(t) = F_0 + F_1 e^{i\omega_1 t} + F_2 e^{i \omega_2 t} +\mbox{h.c.}, \mbox{with:}\vspace{2mm}\\
F_i = \frac{1}{2\sqrt{2}} 
  \left[ \begin{array}{cc}
   f^r_1+i f_1^i & f^r_2+i f_2^i  \\
   f^r_3+i f_3^i  & f^r_4+i f_4^i 
  \end{array}  \right].

\end{array}
 \label{eq:ComplexHam}
\end{equation}
$H(t)$ supports various topological realizations, including the temporal BHZ model as a possible outcome.  Parameters are chosen from $P(f_j^{r,i}) = e^{-(f_j^{r,i}-c)^2/2 \sigma^2}$. We examine the results of Fig.~\ref{Fig:CP}, where we plot $P(\bar{E})$ for $\sigma=1$ and $N=10^4$ realizations, with $\omega_1/\omega_2=2/3$ as well, and $\omega_2=1$.

Once more, a transition is observed. Below $\ccr=2$ we observe a Laplacian distribution:
\begin{equation}
    P(\bar{E})\approx \frac{1}{\beta}e^{-\bar{E}/\beta}\,\,\,\,\,\,[\ccr<2 ]
\end{equation}
[see Fig.\ref{Fig:CP}-a) for $c=1$], while above, we observe a Gaussian:
\begin{equation}
P(\bar{E})\propto e^{-\bar{E}^2/2\sigma_0^2}\,\,\,\,\,[\ccr>2]
\end{equation}
[see Fig.\ref{Fig:CP}-b) for $c=3$]. For a given $\sigma=1$, the standard deviation $\sigma_0$ of $P(\bar{E})$ is an increasing function of $c$, while the normalized distribution $P(\bar{E}/\norm{F_i})$ is described by a standard deviation $\sigma_n$ with a weak dependence on $c$, Fig.\ref{Fig:CP}-c).  For $c=0$ and $\sigma \ll 1$, both $\sigma_0$ and $\sigma_n$ scale as $\sigma_0 \propto \sigma^4$, illustrated in Fig.\ref{Fig:CP}-d). Finally, the nearest neighbor spacing distribution $P(\varepsilon_s)$ shown in Fig.~\ref{Fig:CP}-e) exhibits quadratic level repulsion at small spacings, $P(\varepsilon_s)\propto \varepsilon_s^2 e^{-b \varepsilon_s^2}$, and resembles the spacing distribution of a Gaussian unitary ensemble, for which it holds $b=4/\pi$ \cite{GUHR1998189}. Here we find $b=0.165$ and note that the transition at $\ccr = 2$ is not evident in $P(\varepsilon_s)$.

A particular realization of the model \eqref{eq:ComplexHam} is the random temporal BHZ model 
\begin{align}
H(t)&=\eta_0 \sigma_z +\eta_1^x \sin(\omega_1 t +\phi_1)\sigma_x - \eta_1^z \cos(\omega_1 t +\phi_1) \sigma_z \nonumber \\
&+\eta_2^x \sin(\omega_2 t +\phi_2)\sigma_y - \eta_2^z \cos(\omega_2 t +\phi_2) \sigma_z \,. 
\label{eq:RandomBHZ}
\end{align}
Depending on the parameters, the considered model could belong in a topological dynamical class with $\vert C\vert =1$. For $\eta^j_i=\eta$, $\eta_0= \eta m$ and $\vert m \vert <2$ (topological class), it is well established that the energy transfer is quantized in the near-adiabatic limit $\eta \gg \omega_i$ \cite{PhysRevX.7.041008}. Away from this limit, strong fluctuations are induced by the nonadiabatic driving conditions, for both rationally and irrationally-related frequencies, while the former exhibit more efficient pumping that exceeds the quantized rate (see Fig.~\ref{Fig:BHZComVsIn} of Appendix \ref{app:QuantinzedPumping}). In this regime, the topological properties become less important and the temporal BHZ model is one realization of an ensemble of many that belong to the same symmetry class, making the statistical description of the pumping effect necessary. Only in the strong-drive limit does the physics related to the topological class becomes dominant and quantized energy transfer is restored. 

\section{Parameter Importance} \label{sec:MachineLearning}

A Random-Floquet Hamiltonian approach can be utilized to investigate frequency conversion processes in a relatively large parameter space. Yet there are several questions that are difficult to settle, including the importance of Hamiltonian parameters in resulting a conversion process with high efficiency. To this end, we propose a feature extraction classification algorithm applied to the random temporal BHZ model of Eq.~\eqref{eq:RandomBHZ}, in order to recognize the relevance of topology in nonadiabatic pumps. Machine-learning approaches have been successfully applied in diverse fields including the identification of quantum phases \cite{PhysRevResearch.2.023266,2103.15855}, ab initio solution of many-electron systems \cite{PhysRevResearch.2.033429},  estimation of magnetic Hamiltonian parameters \cite{PhysRevB.99.174426,Kwoneabb0872}, and others. 

 \begin{figure}[t]
	\centering
	\includegraphics[width=1\linewidth]{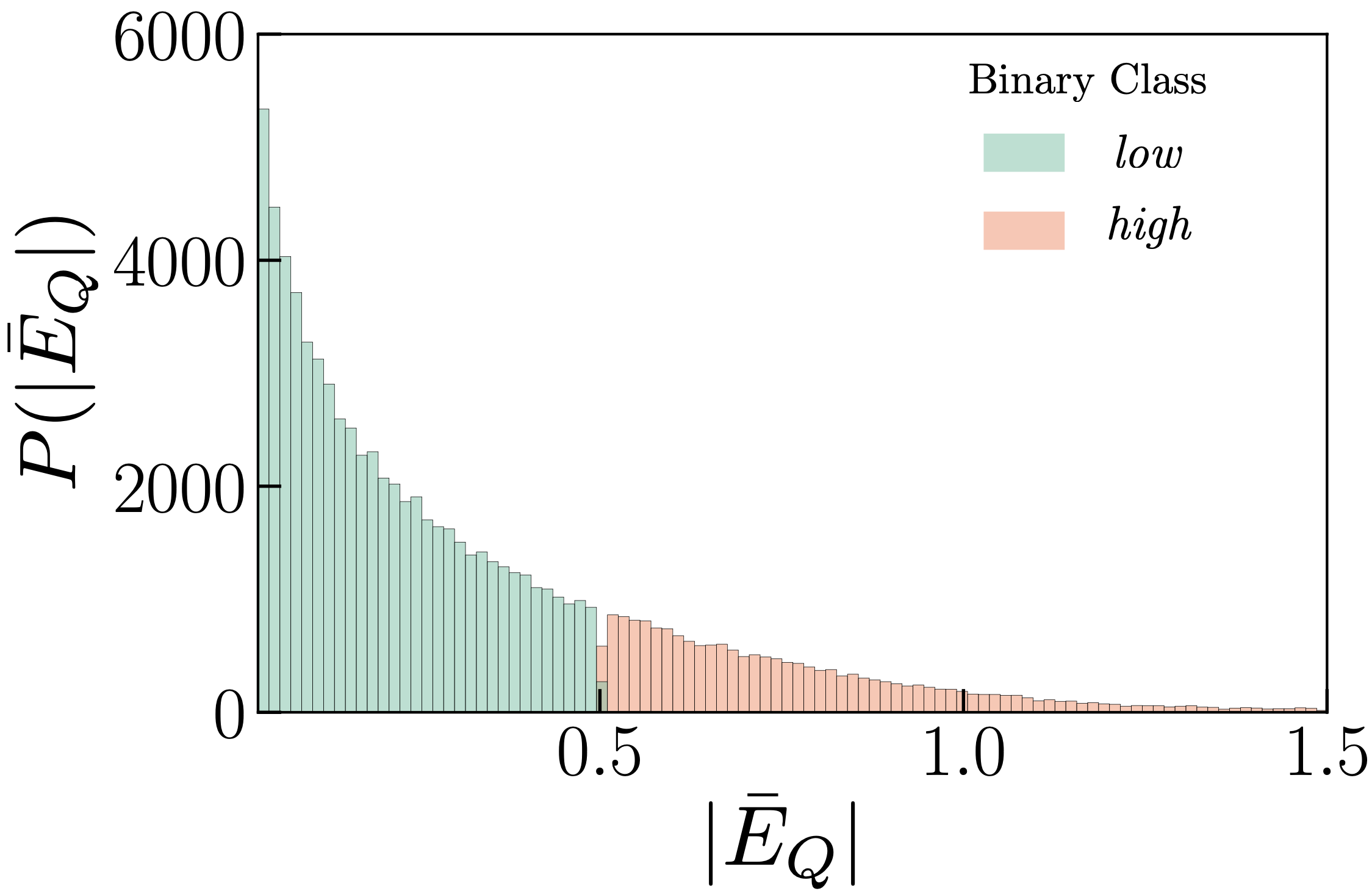}
	\caption{Normalized energy pumping efficiency distribution $P(\vert \bar{E}_Q \vert)\propto e^{-\vert \bar{E}_Q \vert /\beta}$ with $\beta=0.26$ for a binary machine learning classifier. Data are classified into a \textit{low} and \textit{high} efficiency class with a decision boundary at $\bar{E}_d=0.5$.}
	\label{Fig:Binary_Class}
\end{figure}

We generate a dataset of $N=10^5$ elements for binary classification with eight uncorrelated features ($\eta_i^j$, $\phi_i$, and $C$), depicted in Fig.~\ref{Fig:Binary_Class}. Input data with an efficiency below the decision boundary $\bar{E}_Q \leq \bar{E}_d$ correspond to class \textit{low}, while the rest are classified as \textit{high}, with $\bar{E}_Q=\bar{E}/E_Q$ being the normalized efficiency. Parameters $\eta_i^j$ are sampled from a Gaussian distribution with $c=1$ and $\sigma=1$, and phases $\phi_i$ from a uniform in the interval $[0,2\pi$]. The statistical properties of the dataset, together with details on the classification model are given in the Appendix \ref{app:featuresalgo}. Choosing $\bar{E}_d=0.5$, $81\%$ of all data are classified as \textit{low}, while $60\%$ are characterized by a vanishing Chern number $C=0$.  Interestingly, $34\%$ of data with \textit{low} efficiency belong to the $C=1$ class, while it rises to $66\%$ for the $high$ class, indicating that topological models have a higher representation in the \textit{high} class. As features and targets are interacting nonlinearly we employ a gradient-boosted trees model classifier, which exhibits the best performance in terms of commonly used metrics such as recall, precision, and accuracy \cite{10.5555/2815535}.   

A key step towards Hamiltonian engineering targeting high conversion efficiency is understanding the influence of individual features. The relative feature importance $\mc{I}$ reflects how often a feature is used in the split points of a decision tree. We identify $\mc{I}(\eta_0) =24.1\%$, $\mc{I}(\eta^z_2) =15.9\%$, $\mc{I}(\eta^z_1) =13.6\%$, $\mc{I}(\eta^x_1) \approx 12.6\% \approx \mc{I}(\eta^x_1)$, $\mc{I}(\phi_i)=10.6\%$, and $\mc{I}(C) =0.01\%$. Our findings suggest that the uniform $z$ component of the magnetic field is most valuable in achieving high conversion efficiency in nontopological pumps, while an instantaneous topological model is almost irrelevant. The relative feature importance remains unchanged for different decision boundaries $\bar{E}_d \in [0.4,0.6]$ and under the inclusion of a multi-class approach (e.g splitting the data into \textit{low}, \textit{intermediate}, and \textit{high} efficiency).

\section{Analytical bounds on the pumping of arbitrary states} \label{sec:AnalyticalBounds}

Before closing we consider the energy pumping in a doubly driven system which is not initialized in a Flouqet eigenstate. Non-adiabatic aspects of the energy pumping make the pumped power in an arbitrary superposition of Floquet states differ from the simple weighted sum of the pumping rate in each of the eigenstates. To study this, let's look at the work operator
\begin{equation}
    \hat{W}_k(T)=i\omega_k\frac{\partial}{\partial \phi_k}\log U(T)=\omega_k T\frac{\partial}{\partial \phi_k} (\sum\limits_i\varepsilon_i |i\rangle\langle i|)  \,.
\end{equation}

A straightforward manipulation leads to the following expression 
\begin{equation}
    \hat{W}_k(T)=\omega_k T\left(\sum\limits_i\frac{\partial\varepsilon_i}{\partial \phi_k}|i\rangle\langle i| +\sum\limits_{i\neq j} \left(\varepsilon_i-\varepsilon_j\right) |j\rangle \langle i|\partial_{\phi_k} j\rangle\langle i|\right). 
\end{equation}
with $|\partial_{\phi} j\rangle=\frac{\partial}{\partial_{\phi}} |j\rangle$, and $\varepsilon_i$ the Floquet quasi-energies. 

We can separate the work operator into two cases: 2-level systems and larger systems. For 2-level systems, we further assume that the Flqouet quasienergies appear in same-magnitude pairs, $\varepsilon_1=-\varepsilon_2$. By squaring the operator $W_k(T)$ we obtain:
\begin{equation}
W^2=\frac{1}{2}Tr W^2=(\omega_k T)^2\left(\left(\frac{\partial \varepsilon_1}{\partial\phi_k}\right)^2+4\varepsilon_1^2 \left| \langle 1|\partial_{\phi_k} 2\rangle\right|^2\right)
\end{equation}
On the other hand, for multilevel systems, we obtain a lower bound on the energy pumped:
\begin{equation}
\begin{array}c
W^2\ge \frac{1}{N}Tr W^2=\frac{1}{N}(\omega_k T)^2\left(\sum\limits_{i=1}^{N} \left(\frac{\partial \varepsilon_i}{\partial\phi_k}\right)^2\right.\\\left.
+\sum\limits_{i=1,j}^{N}(\varepsilon_i-\varepsilon_j)^2 \left| \langle i|\partial_{\phi_k} j\rangle\right|^2\right)
\end{array}
\end{equation}
Interestingly, the maximum work transferred in a  multidrive system depends on the quantum metric based on changes in the relative phase of the drive. We intend to investigate the maximum work done in future work. 

\section{Discussion} \label{sec:Discussion}

To conclude, we investigated the Floquet statistics for an ensemble of doubly-driven random Hamiltonians in large parameter space, with an emphasis on the distribution of the energy pumping efficiency, by leveraging ideas from RMT.  For nonadiabatic pumps, it holds $P(\bar{E}) = f(\bar{E}/\beta)$ with $\beta \propto \sigma^4$ and $\sigma$ a  scale parameter specific to particular Hamiltonian ensemble. 

Our main finding is that in the spherical ensemble, Sec. \ref{SE-sec}, as well as in the complex Gaussian ensemble, Sec. \ref{CG-sec}, there is a transition in the type of distribution that describes the pupming rate. This transition 
cannot be associated with a symmetry breaking of the instantaneous Hamiltonian. It occurs when the dirve angular frequency and amplitude are of about the same order. While this transition is not associated with the Chern number of an underlying 2d band structure,  it is likely driven by Berry curvature effects which become significant at the same range.

The scaling of energy pumping at small amplitudes requires some additional consideration before closing. Particularly we could ask whether any of our results are described by a Magnus expansion at the range of low-normed Hamiltonians ($\eta\ll 1$) \cite{Eckardt_2015,Casas_2001,BLANES2009151}. We find that such an expansion is not sufficient to capture our numerical results.  First, we calculate the low-driving scaling of the quantity $\bar{E}/E_Q\propto \eta^\alpha$ for specific model realizations, presented in detail in the Appendix \ref{app:QuantinzedPumping}. There is appears that for either a generic or a topological nonadiabatic pump,  $\alpha$ does not exhibit universal properties. Also, as we show in Appendix \ref{app-magnus}, the leading powers for the cross-drive work obtained by a fourth-order Magnus expansion for a generic model are $\bar{E}\sim \eta_0\eta_D^4$, with $\eta_0$ the amplitude of the constant term in the Hamiltonian, and $\eta_D$ the scale of the two drives. It thus becomes apparent that the considered regime of intermediate driving amplitudes for which the pumping efficiency becomes significant, $0 \ll \eta < \omega_i$, lies outside of the applicability of the perturbative approach.

Our results can be directly applied in frequency conversion platforms based on single-qubit quantum devices with a deep level of control, making the implementation of various classes of quantum Hamiltonians possible \cite{PhysRevLett.125.160505, PhysRevLett.126.163602}. The considered models offer a simple interpretation of energy and particle pumping of analogous protocols, including Thouless charge pumping in photonic \cite{Cerjan2020}, ultracold fermions \cite{Nakajima2016}, and single-spin \cite{PhysRevLett.120.120501} systems, quantum pumps in quantum dots \cite{Switkes1905}, and photon pumping in cavities \cite{PhysRevB.99.094311,Rakher2010}. 

Energy pumping between different drives of the same system is indeed a generic feature of multi-driven Floquet systems. Nonetheless, little is known about the distribution of such pumping. Numerically, we managed to characterize some swath of models. We expect, however, that this question could motivate a   Floquet Random-Matrix type theory aimed at energy transfer and similar dynamical properties unique to driven systems. 

Future interesting directions for multi-parameter Floquet quantum Hamiltonians include the statistics of the quantum geometric tensor \cite{PhysRevLett.99.095701,PhysRevLett.126.200604}, or more application-oriented approaches, using deep reinforcement learning algorithms aiming at optimizing frequency conversion processes \cite{2009.07876}. 

\section{ACKNOWLEDGMENTS}

We thank Anushya Chandran and Michael Kolodrubetz for useful discussions. C.P. has received funding from the European Union's Horizon 2020 research and innovation program under the Marie Sklodowska-Curie grant agreement No 839004. We are also grateful to the U.S. Department of Energy, Office of Science, Basic Energy Sciences under  Award  de-sc0019166. GR  is  also  grateful  to  the NSF DMR grant number 1839271, as well as ARO MURI grant FA9550-22-1-0339 supported  GR’s  time  commitment  to  the  project  in  equal shares. NSF provided partial support to C.P. This work was performed in part at Aspen Center for Physics, which is supported by National Science Foundation grant PHY-1607611. G.R. is also grateful for support from the Simons Foundation and the Packard Foundation. 

\appendix

\begin{figure}
\centering
\includegraphics[width=1\linewidth]{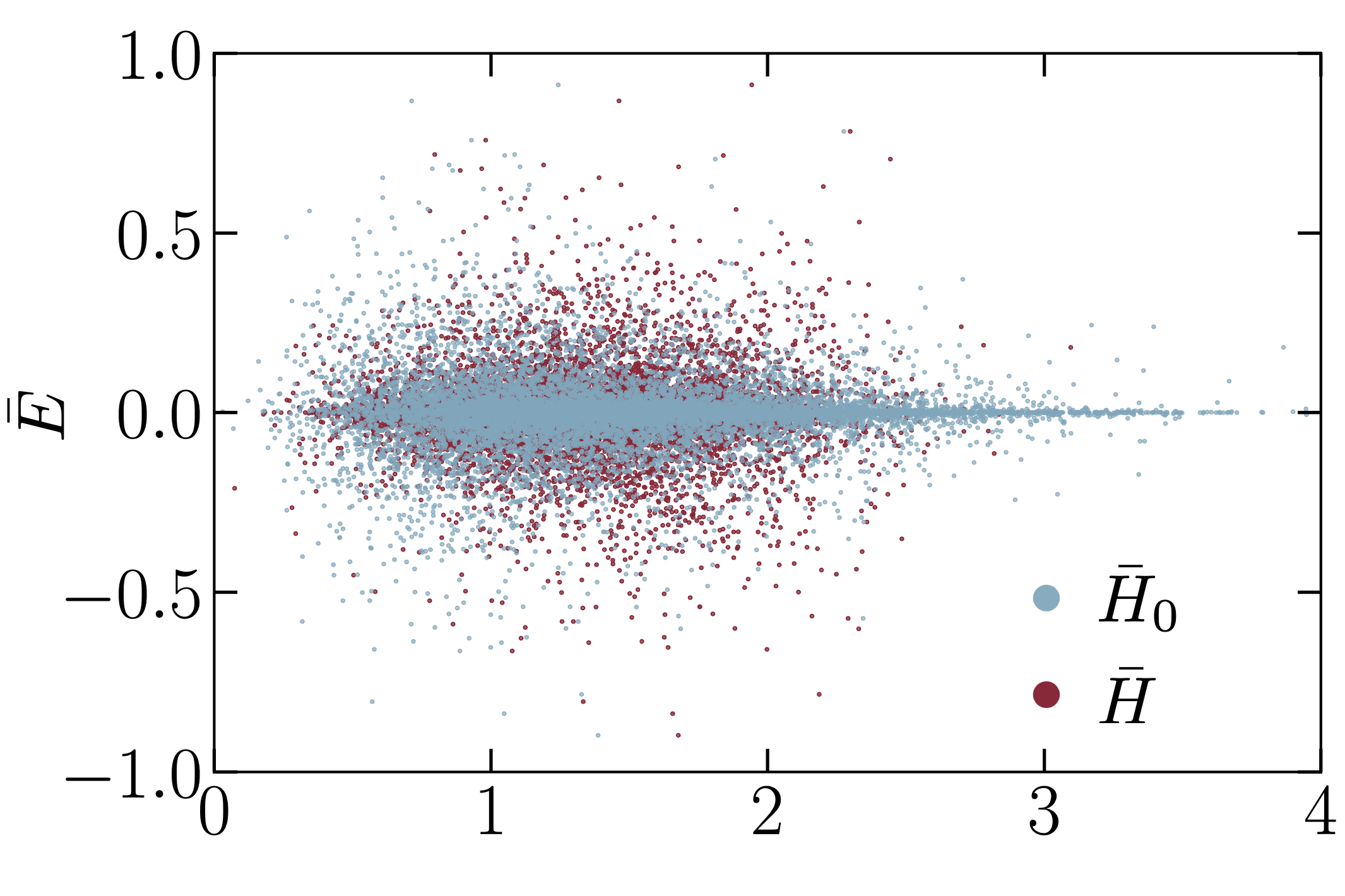}
	\caption{The dependence of $\bar{E}$ on either $\bar{H}_0$ (blue points) or $\bar{H}$ (red points), implying that the two quantities are statistically uncorrelated and no linear relationship between them can be established. Hamiltonian parameters are chosen from a Gaussian distribution with $\sigma=1$ and $c=1$.}
\label{Fig:ScatterPlot}
\end{figure}

\section{Quantized Frequency Conversion} \label{app:QuantinzedPumping}
 Our analysis begins by introducing the temporal analog of the chiral Bernevig-Hughes-Zhang (BHZ) model\cite{Bernevig1757}
 \begin{align}
H(t)=\eta m \sigma_z + H_1(\omega_1 t +\phi_1)+H_2(\omega_2 t +\phi_2) \,, \label{seq:Hamiltonian}
\end{align} 
with $H_1(t)=\eta [\sin(\omega_1 t +\phi_1) \sigma_x -\cos(\omega_1 t +\phi_1) \sigma_z]$, and $H_2(t)=\eta [\sin(\omega_2 t +\phi_2) \sigma_y - \cos(\omega_2 t +\phi_2)\sigma_z]$ the Hamiltonian of the two drives. Here the gap parameter $m$ controls the topological $\vert m \vert<2$ and non-topological $\vert m \vert>2$ regime of the model. The Hall response translates to a quantized pumping of energy between the drives as,
\begin{align}
E_Q= \omega_1 \omega_2 \frac{C}{2\pi} \label{seq:QuantEnergy} \,,
\end{align}
with $C$ the Chern number of the band. Together with the adiabatic requirement $\eta \gg \omega_i$, a necessary condition is that $\omega_1$ and $\omega_2$ are rationally independent, $\omega_1/\omega_2 \equiv \gamma$ with $\gamma \notin \mathbb{Q}$, and the model is in its topological regime $\vert m \vert <2$\cite{PhysRevX.7.041008}. In this respect, energy quantization emerges once the dynamics of the system effectively samples the whole Floquet zone and $C$ takes an integer value. The energy pumping effect for a rational frequency ratio, $\omega_1/\omega_2 \equiv q/p$ for $p,q  \in \mathbb{Z}$, could exceed the quantized value $E_Q$ in the entire topological region and can even be extended in the trivial regime \cite{PhysRevX.7.041008}. In this case, only part of the Berry phase is sampled along a particular periodic path through the Floquet zone, which in turn depends on the choice of the offset phases $\phi_i$. 
\begin{figure}[t]
	\centering
	\includegraphics[width=1\linewidth]{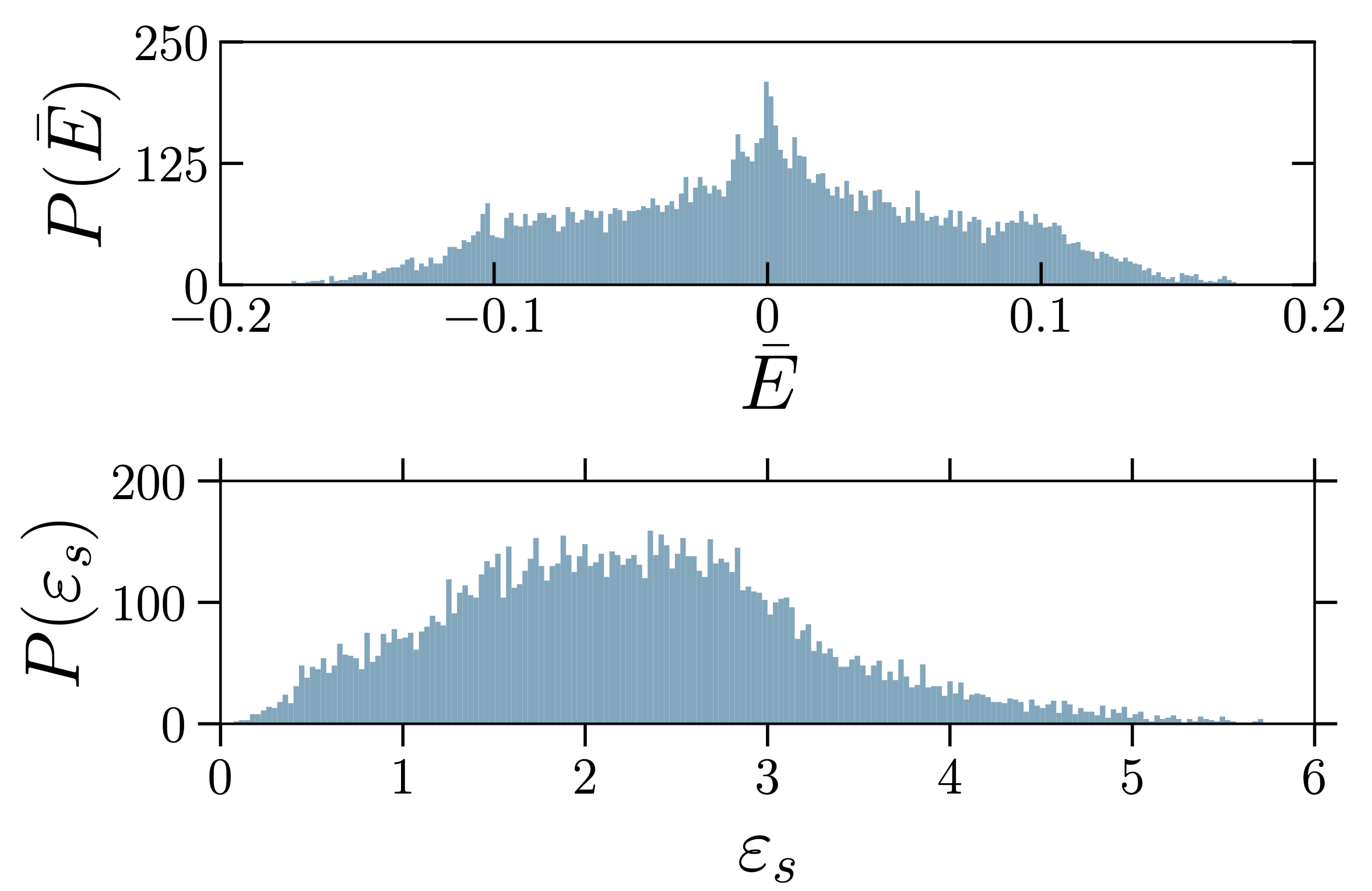}
	\caption{ Energy pumping efficiency distribution $P(\bar{E})$ (upper panel) for a set of Hamiltonian parameters chosen from a spherical ensemble with $\rho=1<\rho_c$ and $N=12000$ realizations. Nearest-neighbor spacing distribution $P(\varepsilon_s)$ (lower panel) for the same values.}
	\label{Fig:UNC1}
\end{figure} 

\begin{figure}[b]
	\centering
	\includegraphics[width=1\linewidth]{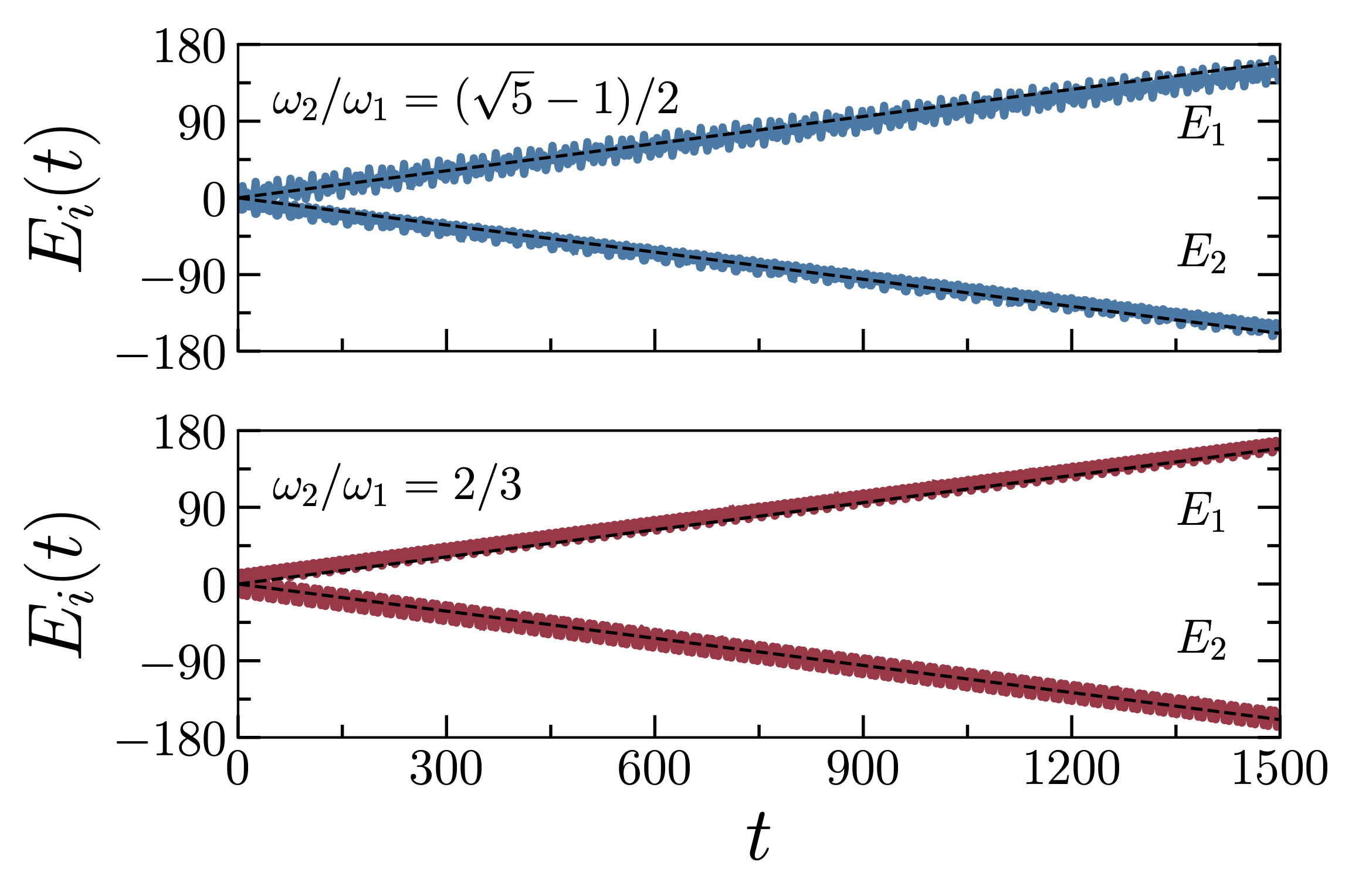}
	\caption{Adiabatic energy flow $E_i(t)$ between the two drives for the temporal BHZ model and incommensurate frequencies $\omega_2/\omega_1=(\sqrt{5}-1)/2$ (upper plane; blue curve) and commensurate frequencies $\omega_2/\omega_1=2/3$ (lower plane; red curve). We choose $\eta=10$ and $m=1.2$. In both cases, the energy flow increases (decreases) at a quantized rate $\bar{E}=E_Q$, illustrated with black dashed lines.}
	\label{Fig:BHZTop}
\end{figure}

When the two frequencies are incommensurate, the energy transfer is maximized when the system is initialized in an eigenstate of $\mathcal{H}_0(t=0)$, while in the opposite case of commensurate frequencies is preferable to consider a Floquet eigenstate initialization, $\vert \psi_0 \rangle =\vert  \psi_F \rangle$ with $U_0(T)\vert  \psi_F \rangle =\varepsilon_F \vert  \psi_F \rangle$. Here $U_0(T)$ is the Floquet single-period evolution operator \cite{Floquet1883}. In Fig.~\ref{Fig:BHZTop} we depict the energy transfer in the adiabatic strong-drive regime $\eta =10$, gap parameter in the topological regime $m=1$, $\omega_1=1$, and $\Delta \phi = \phi_1-\phi_2=0$ for both incommensurate $\omega_2/\omega_1= (\sqrt{5}-1)/2$ (upper panel) and commensurate $\omega_2/\omega_1=2/3$ (lower panel) frequencies. As expected, in both cases the energy pumping rate is $\bar{E}_i = E_Q$. 

\begin{figure}[t]
	\centering
	\includegraphics[width=1\linewidth]{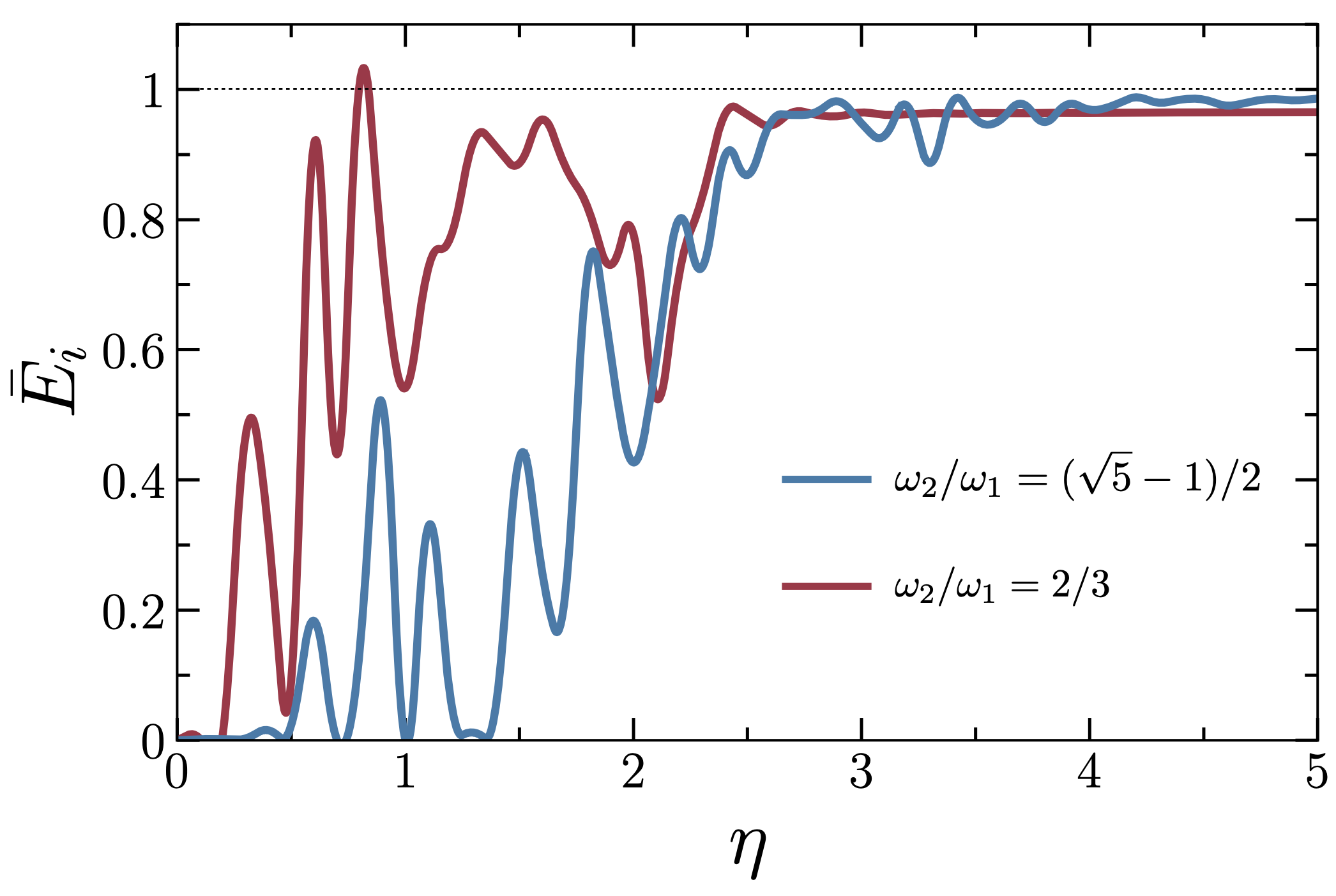}
	\caption{Normalized energy pumping efficiency $\bar{E}_Q=\bar{E}/E_Q$ as a function of the driving amplitude $\eta$ for $m=1$ and two choices of frequency combinations, $\omega_2/\omega_1=2/3$ (red line) and $\omega_2/\omega_1=(\sqrt{5}-1)/2$ (blue line). We note that $\bar{E}_Q$ fluctuates before it converges to unity for sufficiently strong drives $\eta_{\mbox{\tiny ad}} \approx 3$. Fluctuations are stronger for commensurate frequencies, with an efficiency that exceeds the quantized value in the nonadiabatic regime $\eta \leq \omega_i$. }
	\label{Fig:BHZComVsIn}
\end{figure}
The nonadiabatic regime $\eta \ll \omega_i$, in which any bulk gap in the initial Hamiltonian is not maintained under time evolution, is not accessible analytically.  We resort to a numerical calculation of $E_i(t)$ of Eq.~\ref{eq:EnergyBHZ} and subsequent estimation of the normalized conversion efficiency $\bar{E}_Q=\bar{E}/E_Q$, summarized in Fig.~\ref{Fig:BHZComVsIn}. We use $\omega_2/\omega_1=2/3$ (red line) or $\omega_2/\omega_1=( \sqrt{5}-1)/2$ (blue line) and gap parameter $m=1.2$. Strong fluctuations are induced by the nonadiabatic driving conditions, for both rationally and irrationally-related frequencies, while the former exhibit more efficient pumping that exceeds the quantized rate. In this regime, the topological properties become less important and the temporal BHZ model is one realization of an ensemble of many that belong to the same symmetry class, making the statistical description of the pumping effect necessary. Only in the strong-drive limit the physics related to the topological class becomes dominant and quantized energy transfer is restored. 
\begin{figure}[b]
	\centering
	\includegraphics[width=1\linewidth]{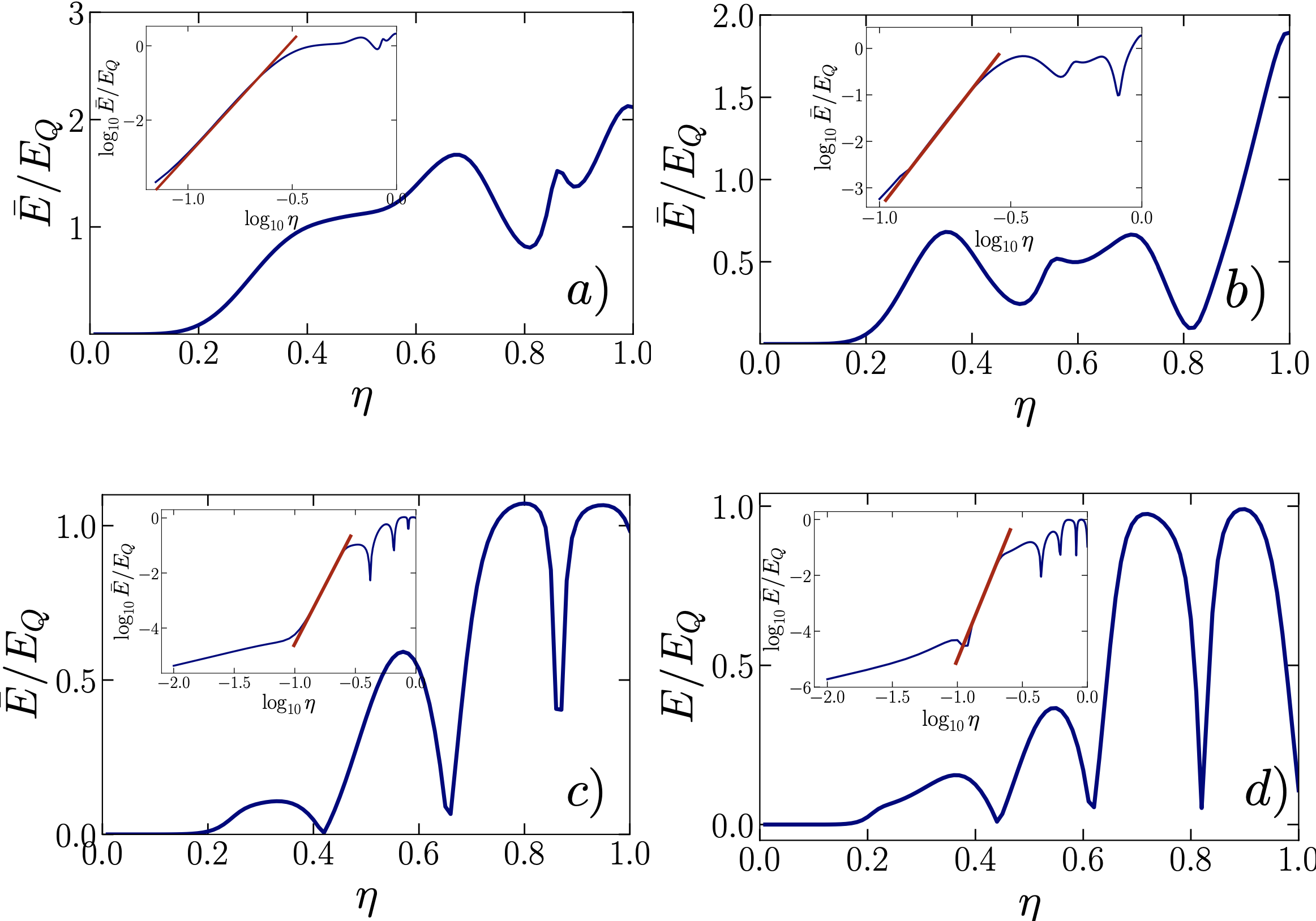}
	\caption{Energy pumping efficiency $\bar{E}/E_Q$ as a function of the driving amplitude $\eta$ in the nonadiabatic limit $\eta < \omega_i$.  a)-b) Energy pumping efficiency for generic model realizations, and c)-d) for topological temporal BHZ model realizations. The insets depict the scaling exponents $\bar{E}/E_Q \propto \eta^{\alpha}$ (red lines) with $\alpha=6$ for a), $\alpha=7$ for b), $\alpha=6$ for c) and $\alpha=12$ for d).}
	\label{Fig:ModelComp}
\end{figure}

\subsection{Low-Driving Expansion} \label{app-magnus}

Here we demonstrate numerically that the low-driving scaling of the quantity $\bar{E}/E_Q$ is model dependent and does not exhibit universal properties.  We consider both the temporal BHZ model Eq.~\eqref{seq:Hamiltonian} as well as a generic Hamiltonian of the form
\begin{align}
H(t)/\eta= F_0 + F_1 \cos(\omega_1 t +\phi_1) + F_2 \cos(\omega_2 t +\phi_2) \,,
\label{eq:Generic}
\end{align}
where $F_i = h_0^{i} \mathbb{1} + h_x^i \sigma_x + h_y^i \sigma_y+h_z^i \sigma_z$.  For all considered models we use $\omega_1/\omega_2 = 2/3$.  We choose two different realizations of \eqref{eq:Generic} for which it holds $\bar{E}/E_Q > 1$ at $\eta =1 $,  with $E_Q= \omega_1 \omega_2/2\pi$, and two realizations of the temporal BHZ model; $m=1.2$,  $\Delta\phi=0.4$ and $m=1$, $\Delta\phi=2.4$.  The overall picture suggested by Fig.~\ref{Fig:ModelComp} is that for low-driving $\eta \ll \omega_i$, the scaling of the quantity $\bar{E}/E_Q \propto \eta ^\alpha$ for either a generic or a topological nonadiabatic pump does not exhibit universal properties. 

A cursory further investigation into the scaling properties of the energy pumping of examples of model 10 reveals why there is no universal scaling. When the scaling of $\bar{E}$ is explored with respect to the magnitude of $F_0$ and separately with respect to the magnitude of $F_1$ or $F_2$, it becomes clear that the scaling observed in Fig. {\ref{Fig:linear_coeff}}
reflects the highly nonlinear nature of frequency pumping. 
Using the very same representative realizations from Fig. \ref{Fig:linear_coeff}, we find that a 4th-order Magnus expansion for the Floquet Hamiltonian yields an even higher scaling power. Within the manifold where $||F_0||=||F_1||=||F_2||$, and $H=\eta_0 F_1 +\eta_D [F_1 \cos (\omega_1 t)+F_2 \cos(\omega_2 t)]$,we find:

\begin{align}
\bar{E}\sim \eta_0 \eta_D^{4} \,.
\end{align}
In the limit of both $\eta_0\ll 1$ and $\eta_D\ll 1$.

\section{Feature Extraction Classification Algorithm} \label{app:featuresalgo}
\begin{figure}[b]
	\centering
	\includegraphics[width=1\linewidth]{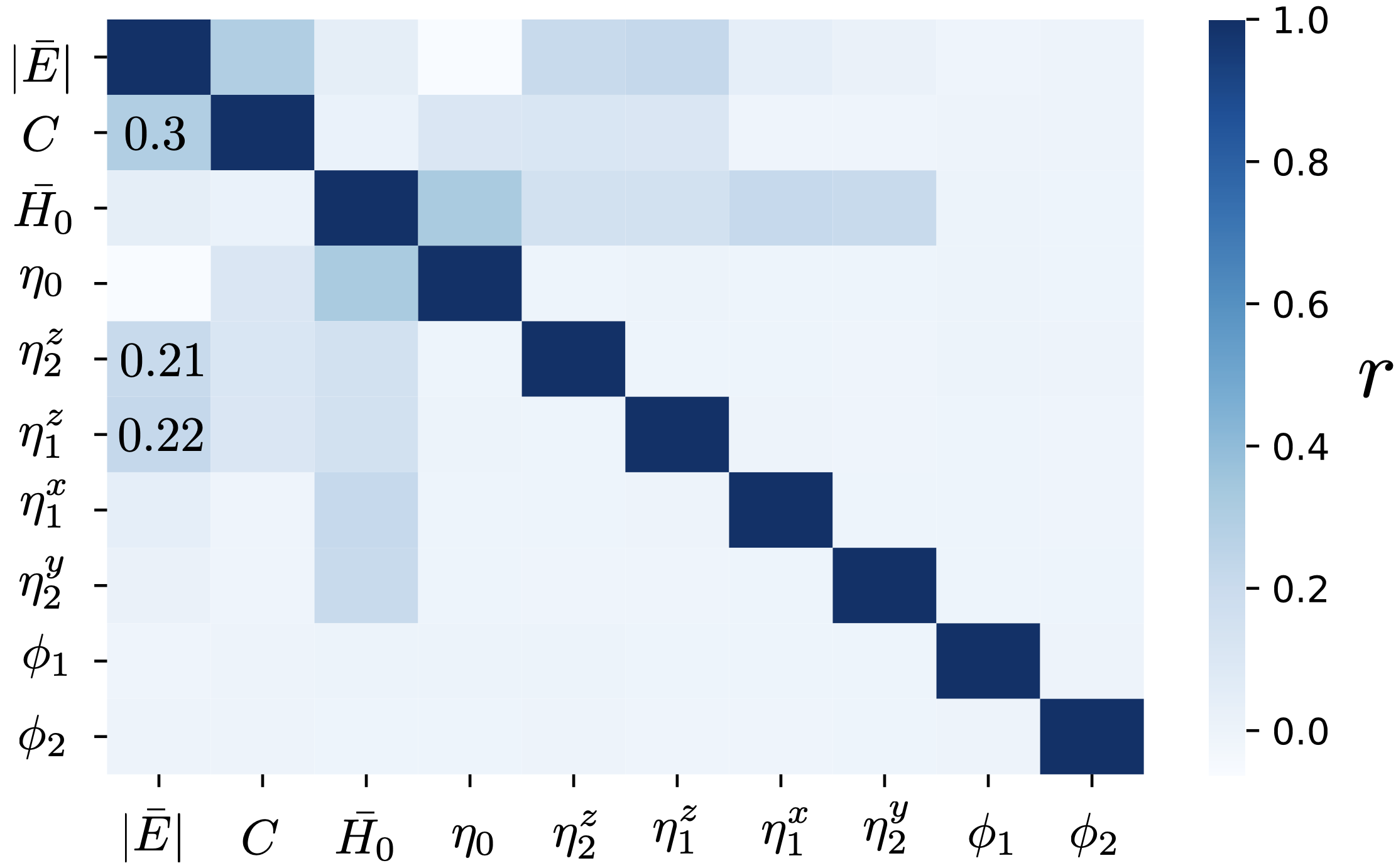}
	\caption{Pearson's correlation coefficient $r$ between pumping efficiency $\bar{E}$, initial time Hamiltonian norm $\bar{H}_0$, driving amplitudes $\eta_{i}^j$, phases $\phi_i$, and Chern number $C$. $\bar{E}$ is only weakly associated with $C$ and $\eta_i^z$, with $r=0.3$ and $r=0.2$ respectively, while $r<0.05$ for the remaining parameters.}
	\label{Fig:linear_coeff}
\end{figure}

In this Appendix, we present details of the classification algorithm employed to derive the importance of individual parameters in resulting in highly efficient energy pumping. 	We consider the following Random-BHZ Hamiltonian 
\begin{align}
H(t)&=\eta_0 \sigma_z +\eta_1^x \sin(\omega_1 t +\phi_1)\sigma_x - \eta_1^z \cos(\omega_1 t +\phi_1) \sigma_z \nonumber \\
&+\eta_2^x \sin(\omega_2 t +\phi_2)\sigma_y - \eta_2^z \cos(\omega_2 t +\phi_2) \sigma_z \,,
\end{align}
and calculate the energy conversion efficiency $\bar{E}$ for commensurate frequencies $\omega_2/\omega_1=3/2$ and Floquet initialization. We generate a dataset of $N = 10^5$ elements, with parameters $\eta_i^j$ sampled from a Gaussian distribution $P(\eta_i^j) =e^{-(\eta_i^j -c)^2/2 \sigma^2}$, with $c=1$ and $\sigma=1$, while phases $\phi_i$ are chosen from a uniform distribution in the interval $[0,2\pi]$. $60\%$ of the generated data belong to the $C=0$ topological class. The energy pumping efficiency distribution decays exponentially $P(\vert \bar{E}\vert)\propto e^{-\vert \bar{E} \vert/\beta}$, with $\beta=0.26$ [see Fig.~\ref{Fig:linear_coeff}]. Maximum efficiency in the ensemble is found at $\bar{E}_{\mbox{\tiny max}}=3.78$, and the mean value at $\bar{E}_{\mbox{\tiny {mean}}}=0.275$. 

We note that the linear correlation between efficiency $\bar{E}$ and Hamiltonian parameters is weak, a result established by calculating the Pearson's correlation coefficient between two datasets $x$ and $y$ of length $N$,
\begin{align}
r_{xy}=\frac{1}{\sigma_x \sigma_y}\sum_{i=1}^N(x_i-\bar{x})(y_i-\bar{y}) \,,
\label{seq:PearsonsCoeff}
\end{align}
where $\bar{x}=\sum_{i=1}^N x_i/N$ the sample mean and $\sigma^2_x=\sum_{i=1}^N(x_i-\bar{x})^2$ the standard deviation. Fig.~\ref{Fig:linear_coeff} presents $r$ between pumping efficiency $\bar{E}$, initial time Hamiltonian norm $\bar{H}_0$, driving amplitudes $\eta_{i}^j$, phases $\phi_i$, and Chern number $C$. $\bar{E}$ is only weakly associated with $C$ and $\eta_i^z$, with $r=0.3$ and $r=0.2$ respectively, while $r<0.05$ for the remaining parameters.

The main issue addressed here is to identify which of the Hamiltonian parameters are important in resulting in high-frequency conversion efficiency, a question treated as a classification problem. We proceed with constructing the model using eight uncorrelated features (amplitudes $\eta_i^j$, phases $\phi_i$ and Chern number $C$). Input data with an efficiency below the decision boundary $\bar{E} \leq \bar{E}_d$ correspond to class \textit{low}, while the rest are classified as \textit{high}. We introduce the normalized pumping efficiency $\bar{E}_Q=\bar{E}/E_Q$ and use the quantized energy transfer $E_Q$ as a reference. This binary classification is visually explained in Fig.~\ref{Fig:Binary_Class} with a decision boundary at $\bar{E}_d=0.5$. $81\%$ of all data belong to the $low$ class (imbalanced data), and $40\%$ of all data are characterized by a finite topological charge $C=1$. In the \textit{low} class, $34\%$ have $C=1$, while in the $high$ class it rises to $66\%$, indicating that topological models have a higher representation in the \textit{high} efficiency class. Mean value of initial time Hamiltonian norm is $\bar{H}_0^{\mbox{\tiny mean}}=1.25$ ($1.33$) for \textit{low} (\textit{high}) class. 

We employ four machine learning classifiers, namely Random Forest, Logistic Regression, Support Vector Machines, and Gradient Boosted Trees, and assess their performance based on commonly used metrics such as $\mc{P}$ precision, $\mc{R}$ recall, and $\mc{A}$ accuracy \cite{10.5555/2815535}. Among them, the Gradient Boosted Trees model classifier can model non-linear interactions between the features and the target and has the highest performance with $\mc{A}=0.96$, $\mc{P}=0.91$ and $\mc{R}=0.88$. The model is trained on $80\%$ of all data and the rest are used as a test set for validation. Class imbalance is treated by adjusting the weight $w$ assigned to each class to $w=1$ for the majority class (\textit{low}) and $w=3$ for the minority class (\textit{high}). The gradient-boosted trees model is a machine learning method that makes predictions by combining a sequence of weak decision tree classifiers based on a gradient-boosting predictive performance \cite{10.5555/2181147}. Once we construct our model, we extract the relative feature importance $\mc{I}$, which reflects how often a feature is used in the split points of a decision tree, averaged over the tree ensemble. We find $\mc{I}(\eta_0) =24.1\%$, $\mc{I}(\eta^z_2) =15.9\%$, $\mc{I}(\eta^z_1) =13.6\%$, $\mc{I}(\eta^x_1) \approx 12.6\% \approx \mc{I}(\eta^x_1)$, $\mc{I}(\phi_i)=10.6\%$, and $\mc{I}(C) =0.01\%$. Our findings suggest that 
the uniform $z$ component of the magnetic field is most valuable in achieving high conversion efficiency and in this limit, the physics related to the topological pumping is less important. 

\begin{figure}[b]
	\centering
	\includegraphics[width=1\linewidth]{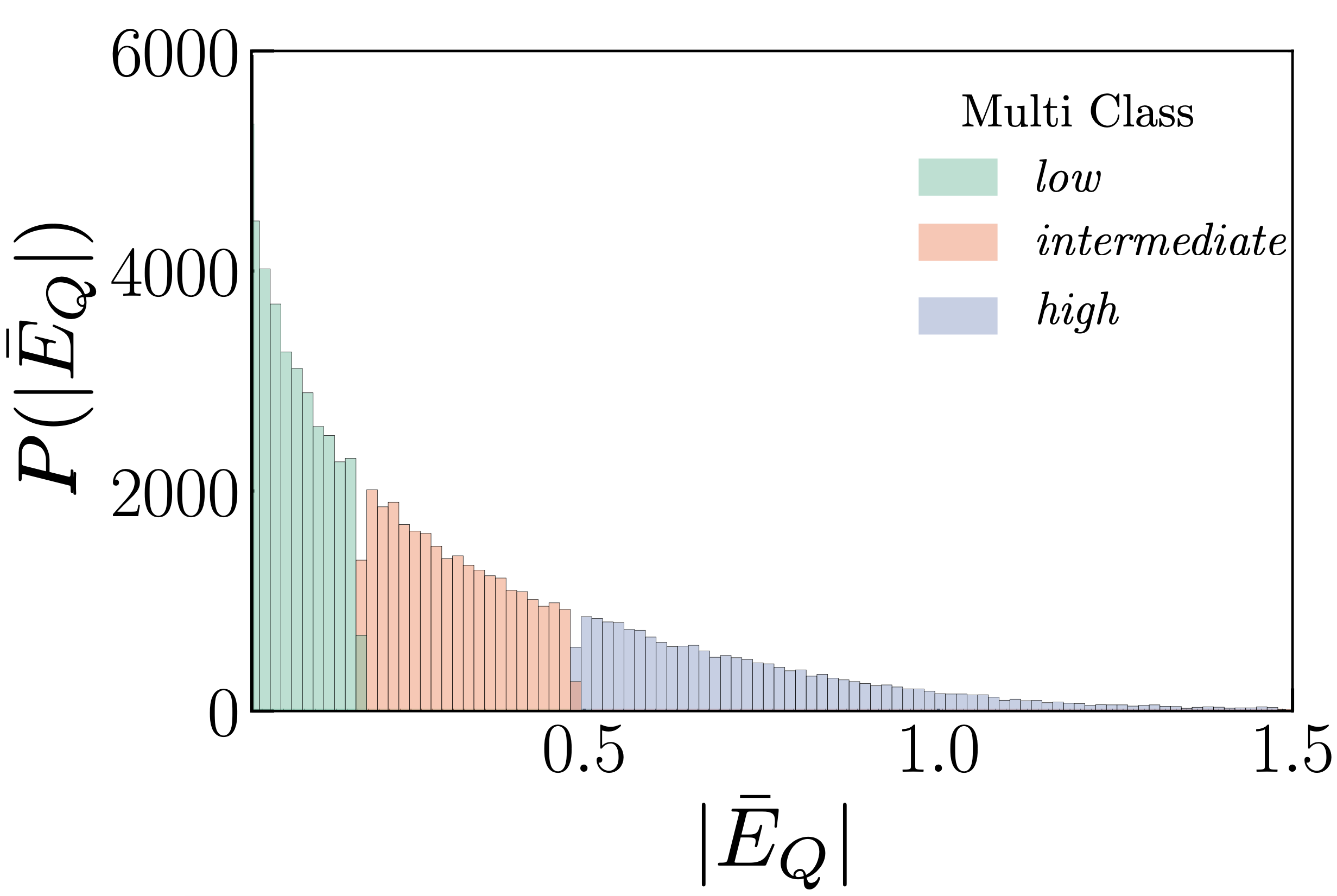}
	\caption{Energy pumping efficiency distribution $P(\vert \bar{E} \vert)$ for a multi-class machine learning classifier. Data are classified into a \textit{low}, \textit{intermediate} and \textit{high} efficiency class with two decision boundaries at $\bar{E}_d=0.2$ and $\bar{E}_d=0.5$.}
	\label{Fig:Multi_Class}
\end{figure}

To complete the description we must also examine the effect of the choices made while constructing the binary classification problem. Since $\bar{E}$ is a continuous variable, we are led to consider difference decision boundaries $\bar{E}_d \in [0.4,0.6]$, and also employ a multi-class approach where data are divided into three classes (\textit{low}, \textit{intermediate}, and \textit{high}), visually explained in Fig.~\ref{Fig:Multi_Class}. In all cases, we arrive at models with similar performances and the same relative feature importance. 
 
 \section{Berry curvature} \label{app:berrycurvature}

\begin{figure*}[]
\centering
\includegraphics[scale=0.29]{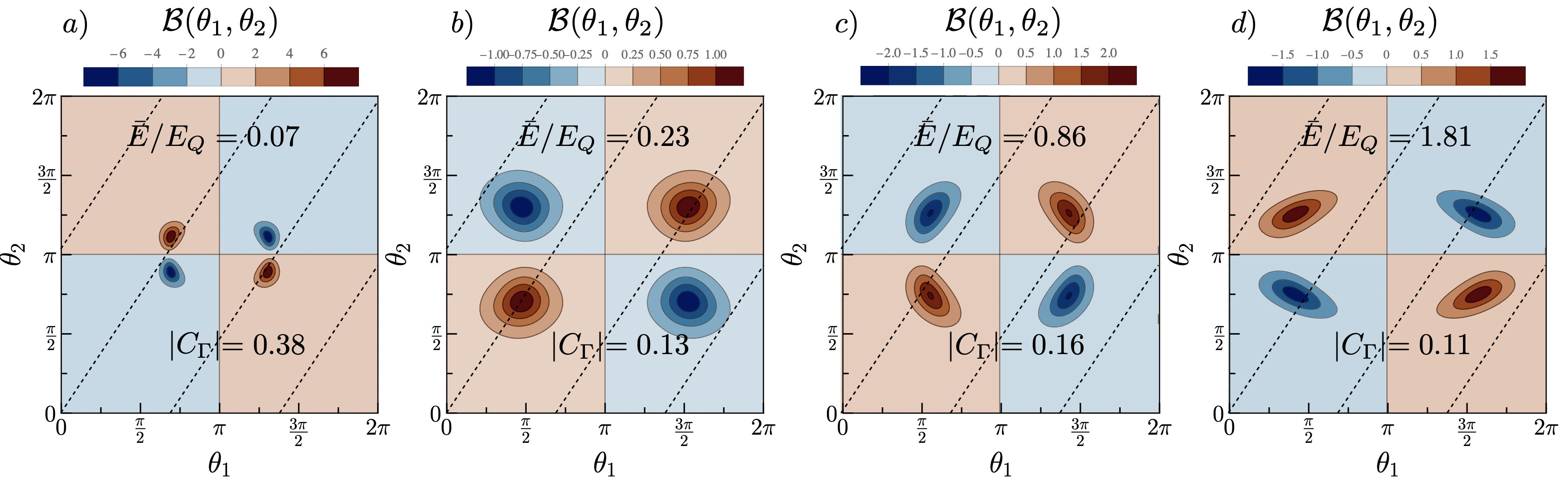}
	\caption{Berry cuvature $\mc{B}(\theta_1,\theta_2)$ of the quasienergy state of Hamiltonian \eqref{eq:Hamiltonian} plotted over the Floquet zone $\theta_i \in [0,2\pi)$ for Hamiltonian parameters chosen from a Gaussian distribution with $c=1$ and $\sigma=1$. For commensurate driving frequencies $\omega_1/\omega_2=2/3$ the system explores a closed periodic path $\Gamma$ through the Floquet zone, drawn by black dashed lines. In all cases, the Chern number of the band is $C=0$.}
\label{Fig:GR_BerryC}
\end{figure*}

\begin{figure*}[]
\centering
\includegraphics[scale=0.29]{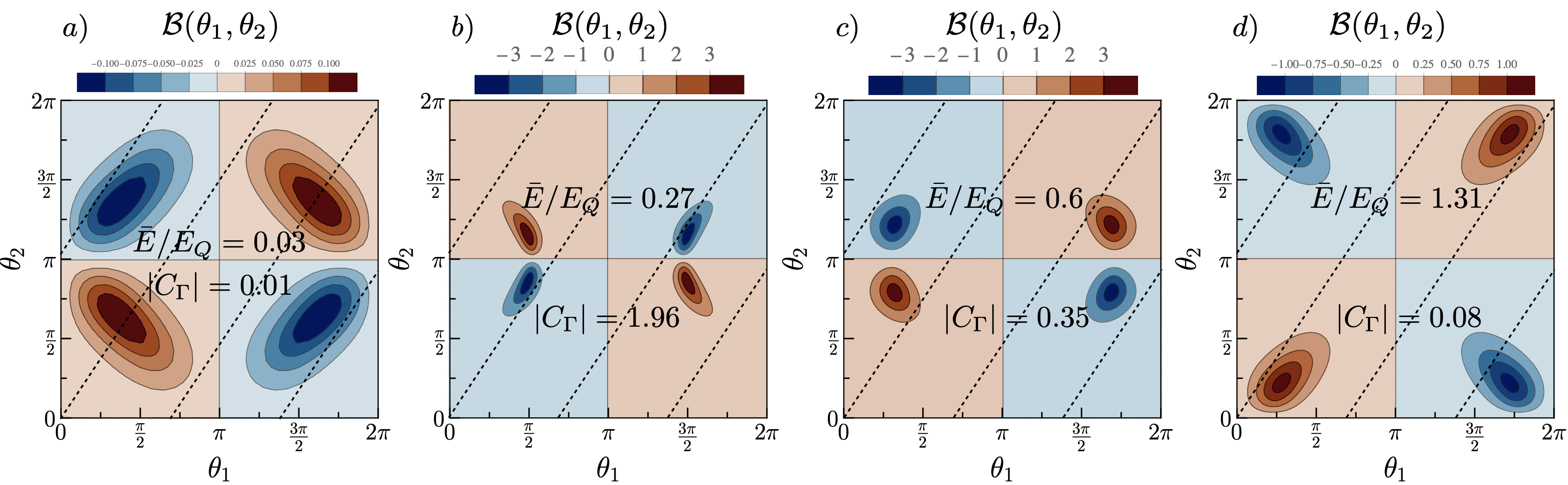}
	\caption{Berry cuvature $\mc{B}(\theta_1,\theta_2)$ of the quasienergy state of Hamiltonian \eqref{eq:Hamiltonian} plotted over the Floquet zone $\theta_i \in [0,2\pi)$ for Hamiltonian parameters chosen from a spherical distribution with $\rho=1.5$. For commensurate driving frequencies $\omega_1/\omega_2=2/3$ the system explores a closed periodic path $\Gamma$ through the Floquet zone, drawn by black dashed lines. In all cases, the Chern number of the band is $C=0$.}
\label{Fig:SP_BerryC}
\end{figure*}

\begin{figure*}[]
\centering
\includegraphics[scale=0.29]{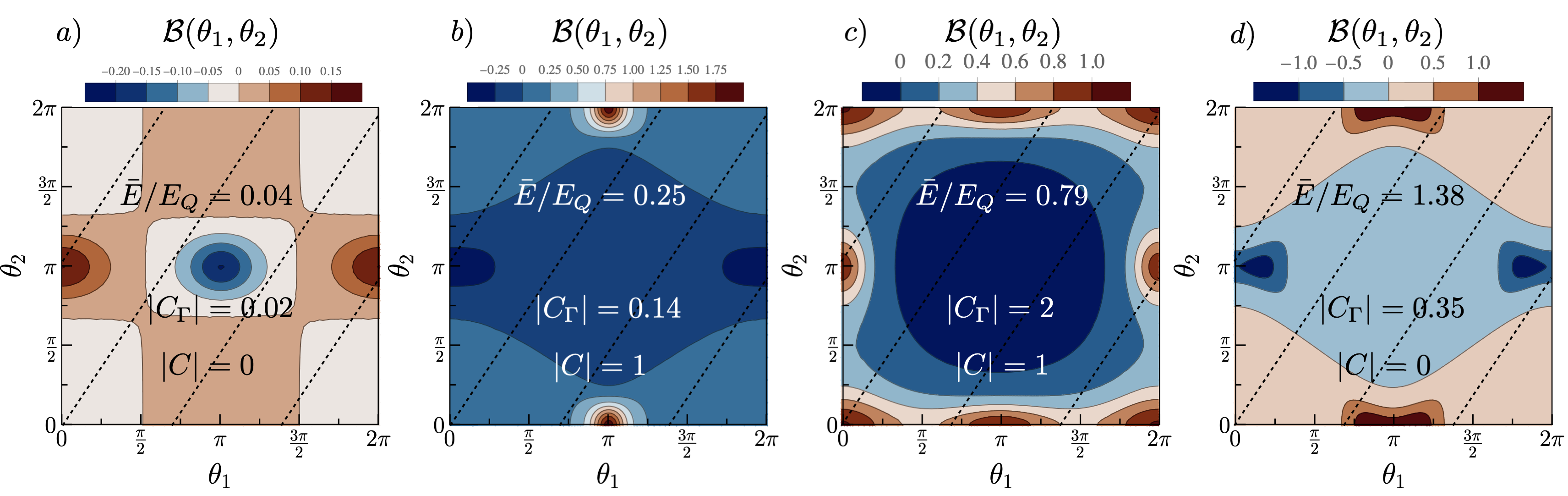}
	\caption{Berry cuvature $\mc{B}(\theta_1,\theta_2)$ of the quasienergy state of the random temporal BHZ Hamiltonian \eqref{eq:EnergyBHZ} plotted over the Floquet zone $\theta_i \in [0,2\pi)$ for Hamiltonian parameters chosen from a Gaussian distribution with $c=1$ and $\sigma=1$. For commensurate driving frequencies $\omega_1/\omega_2=2/3$ the system explores a closed periodic path $\Gamma$ through the Floquet zone, drawn by black dashed lines. The model has several topological realizations with $C=1$.}
\label{Fig:BHZ_BerryC}
\end{figure*}

In this section, we explore geometric aspects of the energy pumping encoded in the Berry curvature of the quasienergy state for various Hamiltonian ensembles. The Berry curvature is defined as follows,
\begin{align}
\mc{B}(\theta_1, \theta_2) &= \langle \pt_{\theta_2} \Psi(\theta_1,\theta_2) \vert \pt_{\theta_1} \Psi(\theta_1,\theta_2) \rangle \nonumber \\ 
&-\langle \pt_{\theta_1} \Psi(\theta_1,\theta_2) \vert \pt_{\theta_2} \Psi(\theta_1,\theta_2) \rangle \,,
\end{align}
where $\theta_i = \omega_i t +\phi_i$ is the phase angle of drive $i$ and the quasienergy state $\Psi(\theta_1,\theta_2)$ is an eigenstate of Hamiltonian $H(\theta_1,\theta_2)$ given in Eq.~\eqref{eq:Hamiltonian}, and use $\omega_1/\omega_2 = 2/3$. In Figs.~\ref{Fig:GR_BerryC}--\ref{Fig:BHZ_BerryC} we present $\mc{B}(\theta_1, \theta_2)$ plotted over the Floquet zone $\theta_i \in [0,2\pi)$ with Hamiltonians parameters chosen from i) a Gaussian ensemble with $c=1$ and $\sigma=1$ (see Fig.\ref{Fig:GR_BerryC}), ii) a spherical ensemble with $\rho=1.5$ (see Fig.\ref{Fig:SP_BerryC}) and for iii) the random temporal BHZ model of Eq.~\eqref{eq:EnergyBHZ} with parameters chosen from a Gaussian with $c=1$ and $\sigma=1$  (see Fig.\ref{Fig:BHZ_BerryC}). The first two models correspond to a linear polarization between the two drives with a vanishing Chern number given by $C=1/(2 \pi) \int_{\mbox{\tiny{FZ}}} d\theta^2 \mc{B}(\theta_1,\theta_2)$. For the random temporal BHZ model, the Chern number can take nonvanishing integer values ($\vert C \vert =1$) depending on the Hamiltonian parameters. 

We note that for commensurate frequencies the system does not sample over the whole Floquet zone, but rather explores a closed periodic path $\Gamma$ depicted by dashed lines in Figs.~\ref{Fig:GR_BerryC}--\ref{Fig:BHZ_BerryC}  \cite{PhysRevX.7.041008,PhysRevB.99.064306}. Within the adiabatic picture, one expects the pumping effect $\bar{E}$ to be roughly the integral of the Berry curvature along the path $\Gamma$, $\bar{E} \propto C_\Gamma \omega_1 \omega_2$, with $C_\Gamma=1/(2 \pi) \int_{\Gamma} d\theta^2 \mc{B}(\theta_1,\theta_2)$. From the results presented in Figs.~\ref{Fig:GR_BerryC}--\ref{Fig:BHZ_BerryC} and explicit calculation of $C_\Gamma$ we conclude that this naive approximation breaks down in the considered nonadiabatic regime and that further effects beyond the local geometrical characteristics of quasienergy states should be taken into account. 

\pagebreak 
\bibliography{RandomMatrix}
 %%%%%%%%%%%%%%%%%%%%%%%%%%%%%
 % Supplemental 
 %%%%%%%%%%%%%%%%%%%%%%%%%%%%%

\end{document}